\documentclass[12pt]{article}
\usepackage{graphicx}
\usepackage{epsfig}
\newcommand{\beq}{\begin{equation}}
\newcommand{\eeq}{\end{equation}}
\newcommand{\beqa}{\begin{eqnarray}}
\newcommand{\eeqa}{\end{eqnarray}}
\newcommand{\beqar}{\begin{eqnarray*}}
\newcommand{\eeqar}{\end{eqnarray*}}

\newcommand{\labell}[1]{\label{#1}} 
\newcommand{\labels}[1]{\label{#1}} 
\newcommand{\reef}[1]{(\ref{#1})}

\newcommand{\eg}{{\it e.g.,}\ }
\newcommand{\ie}{{\it i.e.,}\ }

\newcommand{\norm}[1]{\raise.3ex\hbox{:}#1\raise.3ex\hbox{:}}

\newcommand\prt{\partial}

\newcommand{\gsim}{\mathrel{\raisebox{-.6ex}{$\stackrel{\textstyle>}{\sim}$}}}

\newcommand{\al}{\alpha}

\newcommand{\veps}{\varepsilon}

\newcommand{\eps}{\epsilon}

\newcommand{\Ga}{\Gamma}

\newcommand{\s}{\sigma}

\newcommand\cL{{\cal L}}
\newcommand\cH{{\cal H}}
\newcommand\cR{{\cal R}}
\newcommand\cS{{\cal S}}

\newcommand\tL{{\widetilde L}}
\newcommand\tN{{\widetilde N}}

\newcommand\hp{{\tilde p}}  

\newcommand\hr{{\tilde r}} 

\newcommand{\phd}{\dot{\phi}}

\newcommand{\sph}{{\rm sph}}
\renewcommand{\S}{{\rm S}}
\newcommand{\AdS}{{\rm AdS}}
\newcommand\Pp{P_\phi}
\newcommand\nt{$(n{-}2)$}
\newcommand\mt{$(m{-}2)$}

\begin{document}

\setlength{\unitlength}{1mm}

\thispagestyle{empty}
\rightline{\small hep-th/0008015 \hfill McGill/00-21}
\rightline{\small\hfill BRX TH-472}
\vspace*{2cm}

\begin{center}
{\bf \LARGE SUSY and Goliath}\\
\vspace*{1cm}

Marcus T. Grisaru,\footnote{On leave from Brandeis
University.}$^,$\footnote{E-mail: grisaru@brandeis.edu}
Robert C. Myers\footnote{E-mail: rcm@hep.physics.mcgill.ca}
and
\O yvind Tafjord\footnote{E-mail: tafjord@hep.physics.mcgill.ca}

\vspace*{0.2cm}

{\it Department of Physics, McGill University}\\
{\it Montr\'eal, Qu\'ebec, H3A 2T8, Canada}\\[.5em]

\vspace{2cm} ABSTRACT
\end{center}
We investigate the `giant gravitons' of McGreevy, Susskind and
Toumbas \cite{giant}. We demonstrate that these are BPS configurations
which preserve precisely the same supersymmetries as a `point-like'
graviton. We also show that there exist `dual' giant gravitons
consisting of spherical branes expanding into the AdS component
of the spacetime. Finally, we discuss the realization of the
stringy exclusion principle within this expanded framework.

\vfill \setcounter{page}{0} \setcounter{footnote}{0}
\newpage

\section{Introduction}

One of the intriguing observations to arise in the AdS/CFT correspondence
\cite{juan} -- see also the review \cite{adscft} -- is
the `stringy exclusion principle' \cite{exclus}. In the conformal
field theory, this result is easily understood. A family
of chiral primary operators in the superconformal field theory
terminates at some maximum weight because
the gauge symmetry group has a finite rank. In terms of the dual anti-de 
Sitter
description, these operators are associated with single particle states 
carrying angular
momentum on the internal (spherical) geometry. The appearance of an upper 
bound
on the angular momentum seems mysterious from the point of view of the 
supergravity theory.
Recently, however, McGreevy, Susskind and Toumbas \cite{giant}\ provided an
ingenious mechanism for how the upper bound appears. Rather surprisingly,
their resolution arises through a large distance phenomenon. They suggested
that the supergraviton states expand into the spherical part of the
space-time geometry with a radius proportional to the angular momentum.
The radius of these `giant gravitons' must be smaller than the radius of
the sphere, which through the AdS/CFT is related to the rank of the gauge 
group
in the CFT.  Thus they are able to reproduce precisely the desired upper 
bound
on the angular momentum.

As a consistency check on this picture, it is important that the
energy of this giant graviton state equals the energy of the usual graviton
\cite{giant}. This also suggests that the giant graviton is a
BPS state, preserving some of the supersymmetry. On the other hand, one 
might
think that these spherical brane configurations would break all the
supersymmetries. Antipodal patches of the sphere can be regarded as parallel
portions of branes and anti-branes, so preserving any supersymmetry is 
highly
nontrivial. We will analyze the supersymmetry properties of these
expanded branes, and we show that they indeed are supersymmetric,
and further that the giant graviton preserves precisely the same 
supersymmetries
as a `point-like' graviton.

This supersymmetry analysis thus strengthens the giant graviton
picture, but we will also point out a potential problem with the
interpretation in terms of the stringy exclusion principle. The
problem is the existence of other configurations that could also play
the role of the graviton. In particular, we will show that there
are `dual' giant gravitons, that is, configurations in which a dual
brane expands into the AdS part of the spacetime.
This brane configuration {\it also} has the right
energy and supersymmetries to represent the graviton carrying
a fixed angular momentum. Its
radius is again proportional to the angular momentum on the spherical
part of the spacetime, but because the expansion now occurs in the
non-compact AdS space, there is no upper bound implied.

Hence, instead of a unique candidate for the graviton state, we
have at least {\it three} different ones, including the point-like
gravitons which already arise in the analysis of ref.~\cite{giant}.
Further, since two of the candidates display no upper bound on the angular 
momentum,
there seems to be a problem for the proposed mechanism for the stringy
exclusion principle. Quantum mechanically one might expect
these different states to tunnel into each other, forming a unique
ground state. In this direction, we show that there are
finite-action instanton configurations describing tunneling
between the expanded branes and their zero-size cousins. We suggest
that the exclusion principle might still be realized within
this expanded framework, by speculating that
supersymmetry is spontaneously broken in the regime where there are only
two possible graviton states, {\it i.e.}, when the
angular momentum bound is exceeded.

The paper is organized as follows. We start in Section~\ref{ham} by
defining the various supergravity backgrounds and review the giant
graviton construction in Section~\ref{hama}.
In Section~\ref{hamb}, we then describe the `dual' giant gravitons where
the branes expand into the AdS space, and we follow with an analysis of the
instanton transitions in Section~\ref{hamc}.
Section~\ref{susy} gives an analysis of which supersymmetries
are preserved by the various brane configurations. We conclude in
Section~\ref{discuss} with a discussion of our results and
the possible realization of the exclusion principle.

As this paper was being completed, we were informed about work by
Hashimoto, Hirano and Itzhaki \cite{aki}\ which has a significant
overlap with the present paper.

\section{Expanding Branes} \labels{ham}

We wish to consider various (test) brane configurations in background 
spacetimes
of the form AdS$_{m}\times \S^n$. We will
be considering M2- and M5-branes in the
D=11 supergravity backgrounds with ($m,n$) = (4,7) and (7,4), and also 
D3-branes
in the type IIb supergravity background with ($m,n$) = (5,5). We will give a
general presentation of the brane solutions which encompasses all the three 
cases
simultaneously.

The full line element for the metric on AdS$_{m}\times \S^n$ takes the
form $ds^2=ds^2_\AdS+ds^2_\sph$. We will use global coordinates on $\AdS_m$,
with
\beq
ds^2_{\rm AdS}=-\left(1+{r^2\over \tL^2}\right)dt^2+{dr^2\over
1+{r^2\over \tL^2}}+r^2 d\Omega_{m-2}^2\ ,
\labell{adsmet}
\eeq
and our coordinates on the sphere will be
\beq
ds^2_{\sph}=L^2\left(d\theta^2+\cos^2\theta d\phi^2+\sin^2\theta
d\Omega_{n-2}^2\right)\ .
\labell{sphmet}
\eeq
The radius of curvature for the AdS$_{m}$ is $\tL$, while that for the
$\S^n$ is $L$. For the three cases in which we are
interested,
\beq
(m,n)=\left\lbrace
\matrix{(4,7),&\quad &L=2\tL;\hfill\cr
(5,5),&\quad &L=\tL;\hfill\cr
(7,4),&\quad &L=\tL/2;\cr}\right.
\labell{curvrad}
\eeq
or simply $L={n-3\over2}\tL$
(see, for example, refs.~\cite{duff} and \cite{adscft}).
For the following calculations, a convenient explicit choice of coordinates
for the spherical component of the AdS metric \reef{adsmet} is
\beq
d\Omega_{m-2}^2=d\al_1^2+\sin^2\al_1\left(d\al_2^2+\sin^2\al_2\left(\ \cdots
+\sin^2\al_{m-3}d\al_{m-2}^2\right)\right)\ ,
\labell{Om}
\eeq
and for the $\S^n$ metric \reef{sphmet},
\beq
d\Omega_{n-2}^2=d\chi_1^2+\sin^2\chi_1\left(d\chi_2^2+\sin^2\chi_2\left(\ 
\cdots
+\sin^2\chi_{n-3}d\chi_{n-2}^2\right)\right).
\labell{On}
\eeq

These geometries along with the appropriate form gauge fields, for the three
cases listed in eq.~\reef{curvrad}, comprise maximally supersymmetric 
solutions of the
corresponding supergravity.
The eleven-dimensional supergravity solutions are naturally given in terms 
of the four-form
field strength, which is then proportional to the volume form on
the four-dimensional component of the geometry \cite{freund}.
Alternatively, these solutions can be written in terms of the Hodge-dual 
seven-form
field strength, which is proportional to the volume form on the 
complementary
seven-dimensional part of the space. For the type IIb supergravity solution, 
the
Freund-Rubin ansatz involves choosing the self-dual five-form field strength 
with
terms proportional to the volume form on both of the five-dimensional 
factors
of the full spacetime. The branes in these theories couple to the 
corresponding
potentials. With the coordinates chosen above, we explicitly write
the $(n{-}1)$-form potential on the $\S^n$ as
\beq
A^{(n-1)}_{\phi\chi_1\cdots\chi_{n-2}}=\beta_n L^{n-1} \sin^{n-1}\theta\,
\sin^{n-3}\chi_1\cdots\sin\chi_{n-3}
\equiv \beta_n L^{n-1} \sin^{n-1}\theta\,\sqrt{g_\chi}\ ,
\labell{sphp}
\eeq
where $\sqrt{g_\chi}$ is the volume element on the unit \nt-sphere
described by eq.~\reef{On}. The constant $\beta_n$ is simply a sign: 
$\beta_4=+1=\beta_5$,
while $\beta_7=-1$. 
These sign choices are made so that the four-form field strength
appears with a positive coefficient in both of the M-theory backgrounds.
Given these choices, the $(m{-}1)$-form potential on the AdS part of the space is written
\beq
A^{(m-1)}_{t\alpha_1\cdots\alpha_{m-2}}=-{r^{m-1}\over
\tL}\sin^{m-3}\alpha_1\cdots\sin\alpha_{m-3}\equiv
-{r^{m-1}\over \tL}\sqrt{g_\alpha}\ ,
\labell{adsp}
\eeq
where $\sqrt{g_\al}$ is the volume element on the unit \mt-sphere
described in eq.~\reef{Om}.

\subsection{Giant Gravitons Revisited} \labels{hama}

McGreevy, Susskind and Toumbas \cite{giant} recently
examined branes carrying angular momentum
on the $\S^n$, and they discovered unusual stable configurations in
which an \nt-brane had expanded into the sphere
part of the background geometry. In this section, we review these results.

For all of the cases of interest, the $p$-brane action may be written as
\beqa
S_{p}&=&-T_p\int d^{p+1}\sigma\ \sqrt{-g}+T_p\int P[A^{(p+1)}]
\labell{actp}\\
&=&-T_p\int d^{p+1}\sigma\ 
\sqrt{-g}\left[\,1+{1\over(p+1)!}\,\veps^{i_0\cdots i_{p}}\,
\prt_{i_0}X^{M_0}\cdots\prt_{i_{p}}X^{M_{p}}\,A^{(p+1)}_{M_0\cdots 
M_{p}}\right]
\nonumber
\eeqa
where $g_{ij}$ is the pull-back of the spacetime metric to the
world-volume, \ie
\beq
g_{ij}=\prt_iX^M\prt_jX^N\,G_{MN}\ ,
\labell{pull}
\eeq
and $P[A^{(p+1)}]$ denotes the pull-back of the ($p$+1)-form potential,
which is given explicitly in the second line.
We use $\veps_{i_0\cdots i_p}$ to denote the world-volume volume tensor. 
Hence,
\beq
\veps_{012\cdots p}=\sqrt{-g}\qquad{\rm and}\qquad
\veps^{012\cdots p}=-{1\over\sqrt{-g}}\ .
\labell{volf}
\eeq

In passing, we should comment that the expression in eq.~\reef{actp} is not
the complete world-volume action. For example, we have dropped all of the 
fermions.
It is, of course, consistent with the equations of motion to consider purely
bosonic solutions as we will do in the following --- the fermions will 
reappear
in our discussion of supersymmetry in section \ref{susy}. However, in the 
case
of the M5-brane \cite{5brane,duff} and the D3-brane \cite{dbrane,3brane},
there are additional bosonic world-volume fields: a self-dual
three-form
on the M5-brane, and a conventional gauge field on the D3-brane.
Examining the full equations of motion confirms that for all of the
brane configurations considered here, these additional fields are 
consistently
set to zero. Hence we have verified that it is consistent to work with
the reduced action \reef{actp} for the present analysis.

Following ref.~\cite{giant},
we now wish to find stable test brane solutions where an \nt-brane
has expanded on the $\S^n$ to a sphere of fixed $\theta$ while it orbits
the $\S^n$ in the $\phi$ direction. It is convenient to choose static gauge
in the world-volume theory, where the world-volume coordinates $\s^i$ are 
identified
with the appropriate space-time coordinates:
\beq
\s_0\equiv\tau=t\, ,\ \s_1=\chi_1\, ,\ \ldots\, ,\
\s_{n-2}=\chi_{n-2}\ .\labell{stat}\\
\eeq
We will consider a trial solution of the form
\beq
\theta = {\rm constant}\ ,\qquad\phi=\phi(\tau)\ ,\qquad
r=0\ ,
\labell{try}
\eeq
which corresponds to a spherical \nt-brane of radius $L\sin\theta$
moving around inside the $\S^n$.
From ref.~\cite{giant}\ we know that $\phi$ will be linear in $\tau$,
\ie $\phd=$constant, but we leave $\phi(\tau)$ arbitrary for the purpose
of doing a Hamiltonian analysis below.

For the particular embedding of the \nt-brane in eqs.~\reef{stat}
and \reef{try}, the pullback of the metric \reef{pull} is
\beq
g_{ij} = \pmatrix{
-1+L^2\cos^2 \theta   \dot{\phi}^2&0 \cr
0 & L^2 \sin^2 \theta \,(g_\chi)_{ij}\cr
}\ ,
\labell{pullback}
\eeq
where $(g_\chi)_{ij}$ denotes the metric on the unit \nt-sphere \reef{On}
(with, as in eq.~\reef{stat}, $\chi^i=\s^i$).
Substituting the trial solution \reef{try} into the world-volume action 
\reef{actp} and integrating over the angular coordinates, yields the
following Lagrangian\footnote{Note that we did not remain consistent with
eq.~\reef{sphp} at this point for $n=7$. That is, the sign of the second term
in eq.~\reef{actp} is reversed, which corresponds to the brane having the
opposite charge. In our conventions, the brane expanding into $\S^7$ is
an {\it anti}-M5-brane.}
\beq
\cL_{n-2}=\frac{N}{ L}\left[-\sin^{n-2}\theta\,\sqrt{1-L^2 \cos^2 \theta 
\,\phd^2}+
L\sin^{n-1} \theta \,\phd\right]\ .
\labell{lag2}
\eeq
Here we have introduced the (large positive) integer $N$ using the
quantization of the $n$-form flux on $\S^n$. In each
of the cases of interest, the flux is related to the
tension of the corresponding brane by
\beq
A_{n-2}T_{n-2}={N\over L^{n-1}}\ ,
\labell{quant}
\eeq
where $A_{n-2}$ is the area of a unit \nt-sphere.
The momentum conjugate to $\phi$ becomes
\beq
P_\phi=N\left[\frac{L\sin^{n-2}\theta\cos^2\theta\,\phd}{\sqrt{1-L^2\cos^2\theta
\,\phd^2}}+\sin^{n-1}\theta\right]\ .
\labell{angmom}
\eeq
We can invert this relation to write
\beq
\phd = {1\over L}{p-\sin^{n-1}\theta\over
\cos\theta\,\sqrt{p^2-2p\sin^{n-1}\theta+\sin^{2n-4}\theta}}\ ,
\labell{phidot}
\eeq
where we have introduced $p=P_\phi/N$. The corresponding Hamiltonian
(or Routhian) becomes
\beqa
\cH_{n}&=&P_\phi\phd-\cL_{n-2}\nonumber\\
&=&{N\over L}\sqrt{p^2+\tan^2\theta\,(p-\sin^{n-3}\theta)^2}
\labell{ham2}
\eeqa

Given that the Hamiltonian \reef{ham2} (or the corresponding Lagrangian
\reef{lag2}) is independent of $\phi$, it is clear that equations of motion
will be solved with the momentum (and hence $\phd$) being constant. For
fixed $p$, eq.~\reef{ham2} can be regarded as the  potential that determines
the equilibrium radius of the spherical membrane, \ie fixing the angle 
$\theta$.
In general, one finds
\beq
{\prt \cH_{n}\over\prt\theta}
\propto 
\sin\theta(\sin^{n-3}\theta-p)\left((n-3)\sin^{n-1}\theta-(n-2)\sin^{n-3}\theta+p
\right)\ .
\labell{extrem}
\eeq

\begin{figure}[htb]
\centerline{\includegraphics[width=6.5cm]{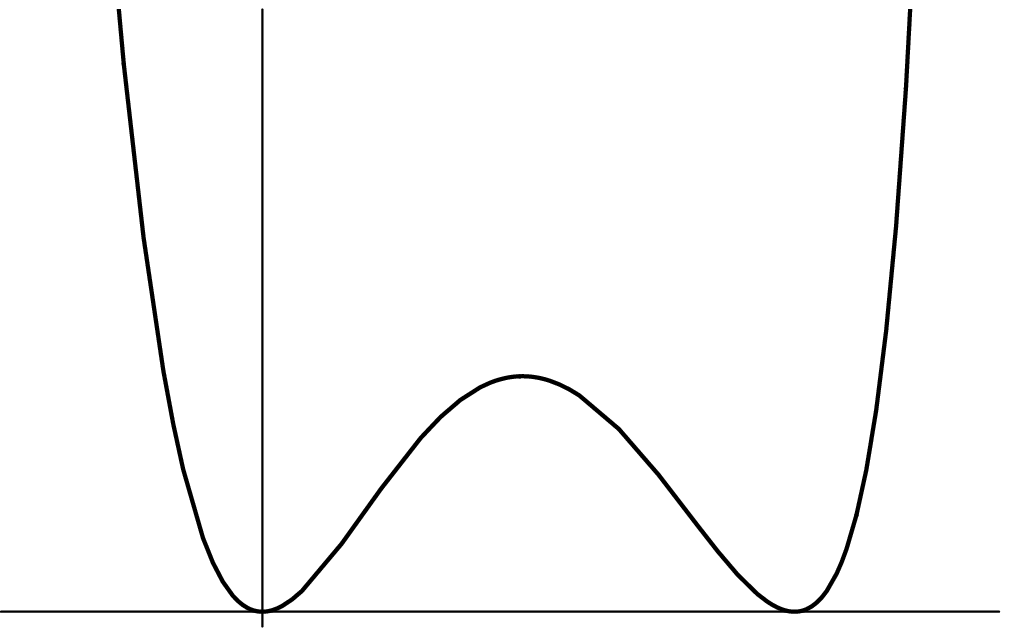}
\hspace{1.5cm}
\includegraphics[width=6.5cm]{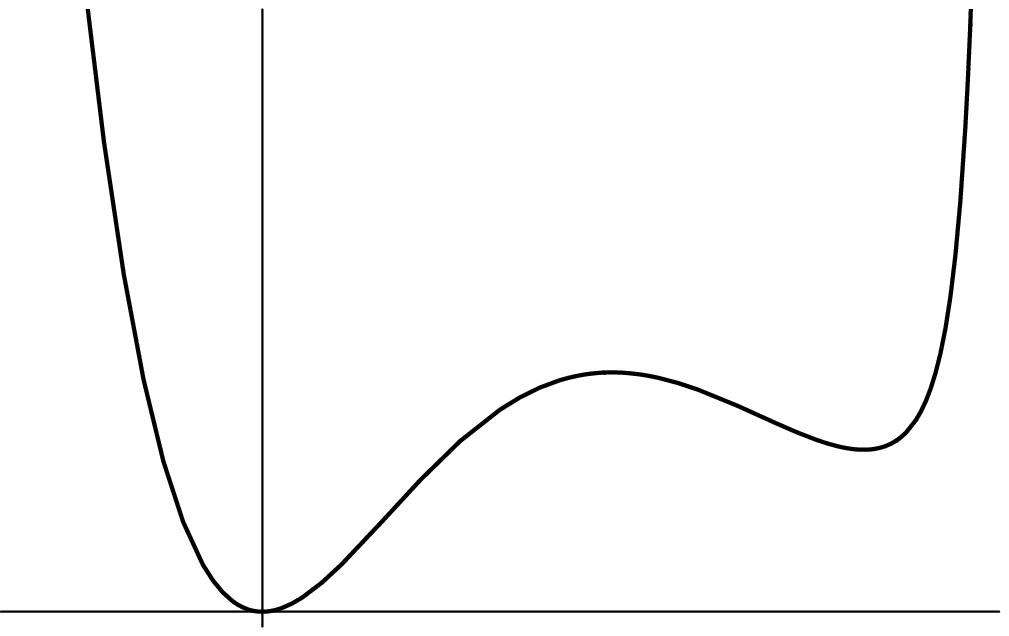}}
\begin{picture}(0,0)
\put(30,50){\small $\cH_n(\theta)$}
\put(76,8){\small $\theta$}
\put(40,38){$p<1$}
\put(113,50){\small $\cH_n(\theta)$}
\put(159,8){\small $\theta$}
\put(123,38){$p>1$}
\end{picture}
\vspace{-1cm}
{
\caption{Energy of expanded \nt-brane, $n$ even,
as a function of its radius. For
$p=P_\phi/N\le 1$ (left figure) there are two degenerate minima. For
$p>1$ the second minimum acquires higher energy and eventually disappears
completely.}\label{fig1}
}
\end{figure}

As illustrated in Figure~\ref{fig1},
for $p\le1$ and $n$ even
there are two
degenerate minima, at $\sin\theta=0$ and $\sin\theta=p^{1/n-3}$,
separated by some
intermediate maximum. At either of the minima, the energy evaluates to
\beq
\cH_{n}={N\over L}p={\Pp\over L}\ .
\labell{hamx}
\eeq
Of course, the minimum at $\sin\theta=p^{1/(n-3)}$ cannot exist
for $p>1$ \cite{giant}. As $p$ grows beyond $p=1$, the minimum at $\theta>0$
first lifts above the one at $\sin\theta=0$, before disappearing
completely for
\beq
p>2\left({n-2\over n-1}\right)^{n-1\over2}\ .
\labell{limp}
\eeq
For the case of interest, $n=4$, this limit corresponds to $p\gsim 1.0887$.

\begin{figure}[htb]
\centerline{\includegraphics[width=6.5cm]{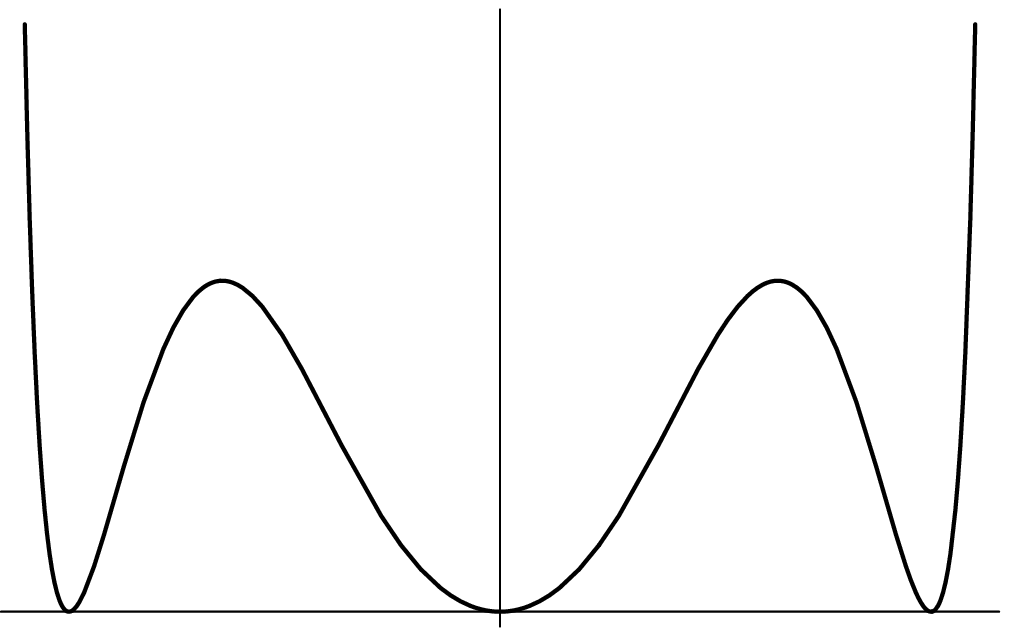}
\hspace{1.5cm}
\includegraphics[width=6.5cm]{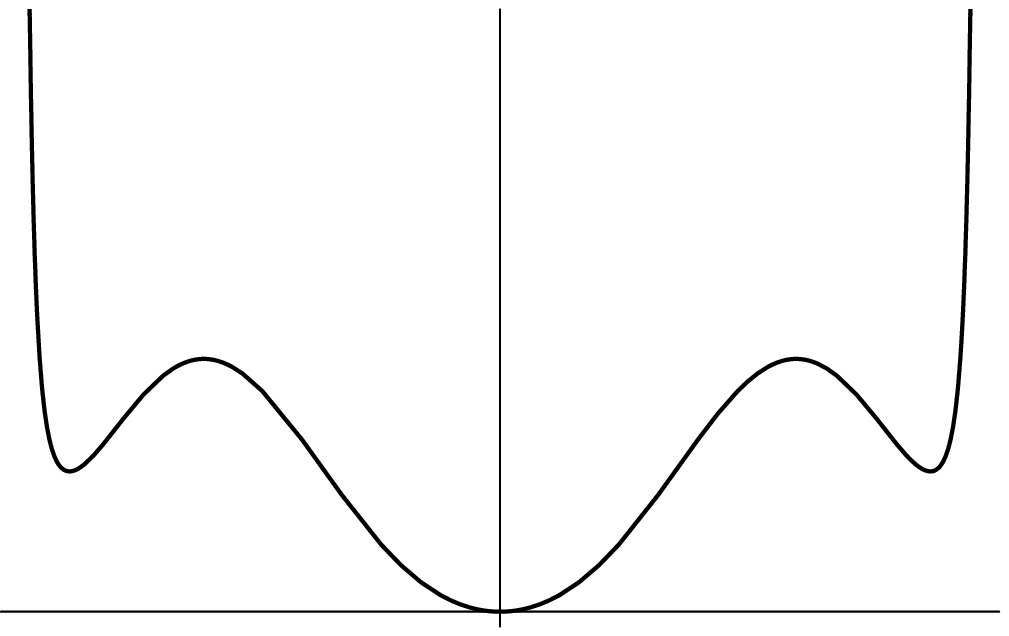}}
\begin{picture}(0,0)
\put(45,50){\small $\cH_n(\theta)$}
\put(76,8){\small $\theta$}
\put(55,38){$p<1$}
\put(128,50){\small $\cH_n(\theta)$}
\put(159,8){\small $\theta$}
\put(138,38){$p>1$}
\end{picture}
\vspace{-1cm}
{
\caption{Energy of expanded \nt-brane, $n$ odd,
as a function of its radius. Same behavior as for $n$ even, except
that the potential is symmetrical.}\label{fig2}
}
\end{figure}

For the case of $n$ odd, the results are similar,
as illustrated in Figure~\ref{fig2}.
For $p\le1$, there are now three
degenerate minima, at $\sin\theta=0$ and at $\sin\theta=\pm p^{1/(n-3)}$, 
separated by some
intermediate maxima. Again at any of the minima, the energy is
$\cH_{n}=\Pp/L$.
As $p$ grows beyond $p=1$, the minima at $\theta\ne0$
are lifted above the one at $\sin\theta=0$ and then disappear
completely if $p$ exceeds the bound given in eq.~\reef{limp}.
For the cases of interest, $n=5$ and 7, this limit corresponds to
$p>1.125$ and $p\gsim 1.1574$, respectively.

We should comment on two related aspects of these results.
First, in our figures
we allow $\theta$ to take negative values. This range, $\theta<0$, 
corresponds
to the \nt-brane expanding into the $n$-sphere, but with the opposite 
orientation
for the angular coordinates. For $n$ even, this is equivalent to an 
anti-brane
with momentum $\Pp$ expanding into the $n$-sphere. For $n$ odd, this is in 
fact
the identical configuration of a brane expanding into the $n$-sphere. This 
explains
a second curious point, namely, for $n$ odd, the potential \reef{ham2} is 
even under
$\theta\rightarrow-\theta$, but there is no such symmetry for $n$ even. In 
the first
case, the configurations with
$\theta<0$ are actually redundant, being physically
equivalent to those with $\theta>0$. In contrast, for $n$ even, positive and 
negative
$\theta$ are distinct physical configurations, and our results show there 
are no
stable expanded configurations for an anti-brane carrying (positive) angular 
momentum $\Pp$.
The latter result also applies for the case of $n$ odd, but requires an 
additional
calculation. To describe an anti-brane, one would consider a test brane with 
the
opposite charge. That is,
one would reverse the sign of the second term in the action
\reef{actp}. Repeating the above analysis in this case results in a 
potential of
the form
\beq
\overline{\cH}_{n}={N\over L}\sqrt{p^2+\tan^2\theta(p+\sin^{n-3}\theta)^2}\ 
.
\labell{ham2a}
\eeq
As expected for $n$ even, this potential is identical to that in 
eq.~\reef{ham2} up to
$\theta\rightarrow-\theta$. On the other hand, for $n$ odd, it is easy to 
see
that the anti-brane potential \reef{ham2a} has no extrema except at 
$\theta=0$.

From the point of view of $m$-dimensional supergravity in the
AdS space, the stable brane configurations correspond to massive states with
$M = P_\phi/L$. The motion on the $\S^n$ means that these states are
also charged under a $U(1)$ subgroup of the $SO(n+1)$ gauge symmetry in
the reduced supergravity theory. With the appropriate normalizations,
the charge is $Q= P_\phi/L$, and hence one finds that these
configurations satisfy the appropriate BPS bound \cite{giant}.
One can therefore anticipate that all of these configurations should
be supersymmetric, and we
confirm this result with an explicit construction of the
residual supersymmetries in Section \ref{susy}.

It is also interesting to consider the motion of these stable 
configurations.
Evaluating eq.~\reef{phidot} for any of the above solutions,
remarkably one finds the same result: $\phd=1/L$, independent of $\Pp$! Note 
then
that the center of mass motion for any of the configurations in the full 
($m$+$n$)-dimensional
background is along a null trajectory, since
\beq
ds^2=-(1-L^2\,\cos^2\theta\,\phd^2)dt^2=0
\labell{null}
\eeq
when evaluated for $\theta=0$ and $\phd=1/L$.
This is, of course, the expected result for a massless `point-like'
graviton, but it applies equally well for the expanded brane configurations.
However, note that in the expanded configurations, the motion
of each element of the sphere is along a timelike trajectory, with
$ds^2=-\sin^2\theta\,dt^2$.

Naively, one might argue that the zero-size solution at $\sin\theta=0$
is unphysical and that the true physical graviton should be
identified with the expanded brane. In particular, one might observe
that if one evaluates the expression \reef{angmom} for the angular
momentum by first taking $\phd=1/L$ and then $\theta=0$, the result
is $\Pp=0$.
Examining more closely, however, it is easy to show that there is a limit
$\phd\rightarrow1/L$ and $\theta\rightarrow0$ such that $\Pp$ remains 
finite.
This is, of course, analogous to the case in ordinary relativistic 
mechanics,
where a limiting procedure (with $m\rightarrow0, v\rightarrow 1$) is 
required
in order to see that zero-mass particles can carry finite momentum. In the
present case, the limit
is slightly unusual in that it involves approaching $\phd=1/L$ from
above. That is,
the center of mass velocity exceeds the speed of light. However,
the motion of the elements of the sphere always remains subluminal!
Hence such a discussion cannot rule out the
point-like configuration as unphysical, and it seems like the
zero-size state needs to be taken into account as well.

\subsection{Giant Gravitons in AdS} \labels{hamb}

In the previous section, we have seen that a spherical \nt-brane 
configuration
has the same quantum numbers as the point-like (super)graviton.
Motivated by the analysis of ref.~\cite{dielec} (see discussion in
Section~4), one might also consider
the possibility of a brane expanding into the AdS part of the spacetime.
In this section, we will show that there is in fact a stable expanded
\mt-brane configuration in the AdS space, which again carries the same
quantum numbers as the point-like graviton.

In this case, we again begin with the same world-volume action \reef{actp}.
Now, however, we wish to find stable solutions where an \mt-brane has 
expanded
into the AdS$_m$ space to a sphere of constant $r$ while it orbits in the 
$\phi$
direction on the $\S^n$. Choosing static gauge, we identify
\beq
\sigma_0 \equiv \tau= t\, , \  
\sigma_1 = \al_1\, , \  \ldots\, , \ \sigma_{m-2} =
\al_{m-2}\ .
\labell{stat2}
\eeq
Our trial solution will be
\beq
\theta = 0\ ,\qquad\phi=\phi(\tau)\ ,\qquad
r={\rm constant}\ .
\labell{try2}
\eeq

\begin{figure}[htb]
\centerline{\includegraphics[width=6.5cm]{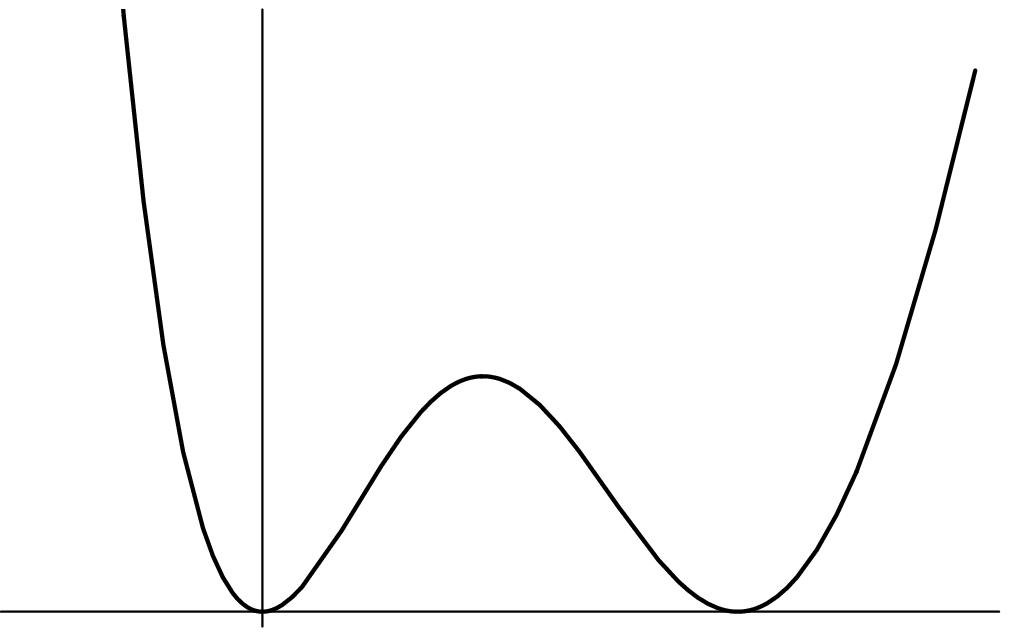}
\hspace{1.5cm}
\includegraphics[width=6.5cm]{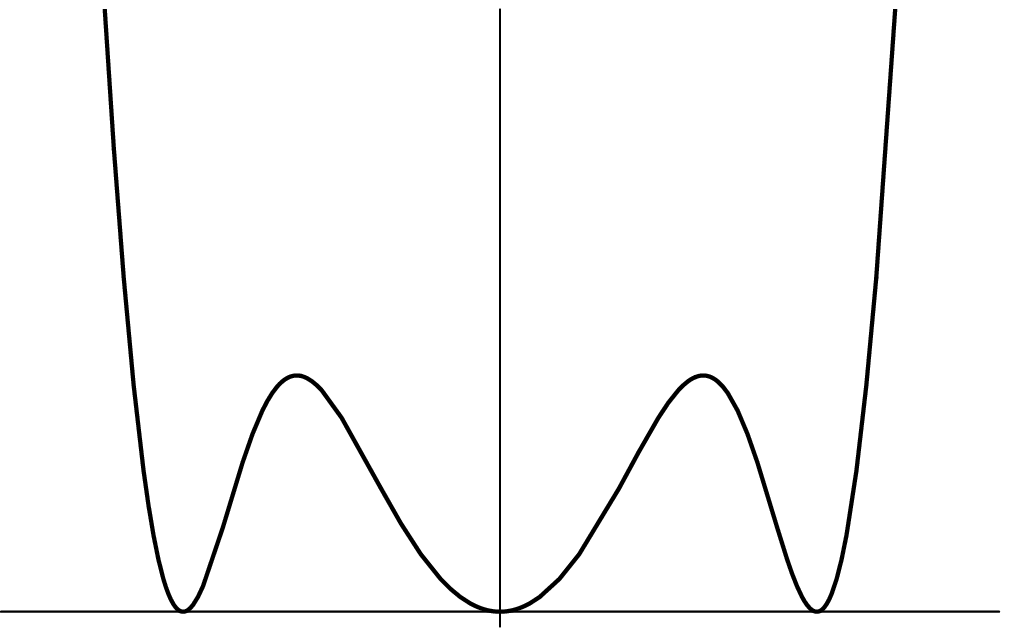}}
\begin{picture}(0,0)
\put(30,50){\small $\cH_m(r)$}
\put(76,8){\small $r$}
\put(40,38){$m$ odd}
\put(128,50){\small $\cH_m(r)$}
\put(159,8){\small $r$}
\put(132,38){$m$ even}
\end{picture}
\vspace{-1cm}
{
\caption{Energy of an \mt-brane expanding into AdS$_m$
as a function of its radius.
}\label{fig3}
}
\end{figure}

Now one can calculate the pull-backs of the metric and the $(n{-}1)$-form
potential, substitute the trial solution and integrate over the
angular directions. The resulting Lagrangian is
\beq
\cL_{m-2}=A_{m-2} T_{m-2}
\left[-r^{m-2}\sqrt{1+{r^2\over \tL^2}-L^2\dot{\phi}^2}
+{r^{m-1}\over \tL}\right],
\labell{clown}
\eeq
where $A_{m-2}$ is the area of a unit $\S^{m-2}$. Note that 
this action actually
describes an {\it anti-brane} by the conventions in eqs.~\reef{adsp} and \reef{actp}
--- that is, the sign of the second term in the action \reef{actp} has been reversed,
corresponding to choosing the opposite brane charge. As we will see,
this choice is
required to produce an expanded brane. As in eq.~\reef{lag2}, it is
convenient to introduce\footnote{Note that $\tN$ is {\it not} the integer
appearing in the background flux quantization condition \reef{quant}
for the M-theory cases $m=4,7$. In terms
of the \mt-brane tension, the latter becomes $A_{m-2}T_{m-2}=
b(m) N^{m-3\over 2}/L\tL^{m-2}$, where $b(4)=\sqrt{2}$ and
$b(5)=1=b(7)$. This follows from the relation between the M2- and
M5-brane tensions, $2\pi T_5=T_2^2$, as well as eqs.~\reef{curvrad}
and \reef{quant}.}
\beq
A_{m-2} T_{m-2}={\tN\over L\,\tL^{m-2}}\ .
\labell{fake}
\eeq
The conjugate momentum for $\phi$ now becomes
\beq
P_\phi=\tN{r^{m-2}\over 
\tL^{m-2}}{L\,\dot{\phi}\over\sqrt{1+{r^2\over\tL^2}-L^2
\dot{\phi}^2}}\ .
\eeq
Introducing the notation $\hp=P_{\phi}/\tN$, one calculates the
Hamiltonian to be
\beq
\cH_{m}=\Pp\phd-\cL_{m-2}
={\tN\over L}\left[
\sqrt{\left(1+{r^2\over \tL^2}\right)\left(\hp^2+{r^{2m-4}\over \tL^{2m-4}}
\right)}-{r^{m-1}\over  \tL^{m-1}}\right]\ .
\labell{green2}
\eeq
Examining $\prt \cH_m/\prt r=0$, one finds minima
located at
\beq
r=0\qquad {\rm and}\qquad\left(r/\tL\right)^{m-3}=\hp\ .
\labell{minix}
\eeq
The energy at each of the minima is $\cH_{m}=\tN\hp/L=P_\phi/L$,
matching the BPS mass found in the previous section.
The potential for even and odd $m$
is illustrated in Figure~\ref{fig3}.
Note that for $m$ even, there is  a single minimum with $r>0$,
while for $m$ odd, there are two such minima, one on either side of $r=0$.
The physical reasons for this structure are the same as in the
case for the branes expanding on the $n$-sphere.
Also as before, one can show that for the case of $m$ odd, there
are no static solutions (other than $r=0$) for an anti-\mt-brane
expanding into the AdS space.
An essential difference from that case, however, is that
the minima corresponding to expanded branes persist for arbitrarily large 
values of $\hp$.

As we did above,
we consider the center of mass motion of these brane configurations.
One again finds for any of the stable minima that $\phd=1/L$, independent
of $\Pp$. The center of mass motion then follows a null trajectory in the
full ($m$+$n$)-dimensional background spacetime for either the point-like
state or the branes that have expanded into the AdS space. In the latter 
case,
the motion of each element of the sphere is along a time-like trajectory
with $ds^2=-(r/\tL)^2\,dt^2$.

\subsection{Instanton Transitions} \label{hamc}

So far we have seen that for a given background $\AdS_m\times\S^n$,
there are three potential brane configurations to describe a graviton
carrying angular momentum $\Pp$: the `point-like' graviton, the
giant graviton of ref.~\cite{giant} consisting of a spherical
\nt-brane expanded out into the $n$-sphere, and a `dual' giant
graviton consisting of a spherical \mt-brane which expands out into
the AdS space.
Quantum mechanically one might expect these three states to mix,
and so one is motivated to look for instanton solutions describing
tunneling from one state to another.
In this section we will derive explicit
expressions for the instantons evolving between the
expanded \nt-brane state and the zero-size state, as well as between
the expanded \mt-brane state and the point-like state. It would be 
interesting
to construct instantons describing tunneling directly between the
two expanded brane configurations, but it is not clear that this can
be done in the framework of test-branes (at least in the case of $\rm{D}=11$).

\begin{figure}[htb]
\centerline{\includegraphics[width=6.5cm]{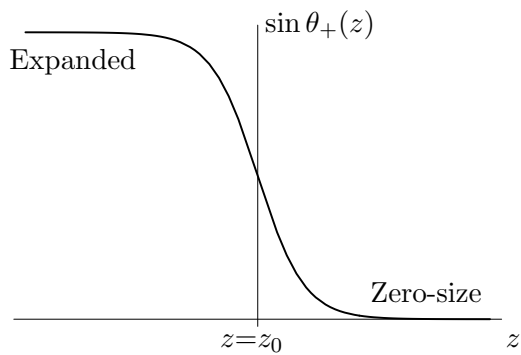}}
\begin{picture}(0,0)
\put(52,45){\small Expanded}
\put(80,8){\small $z{=}z_0$}
\put(100,14){\small Zero-size}
\put(118,8){\small $z$}
\put(86,50){\small $\sin\theta_+(z)$}
\end{picture}
\vspace{-1cm}
{
\caption{Instanton transition from the expanded \nt-brane on $\S^n$ to the
collapsed one (plot made for $m=4$).}\label{fig4}
}
\end{figure}

We start with the instanton for the \nt-brane in $\S^n$.
To begin, we extend the ansatz \reef{try} to allow for
a time dependent angle $\theta(\tau)$.
The Lagrangian (\ref{lag2}) is then extended to
\beq
\cL_{n-2}=\frac{N}{ L}\left[-\sin^{n-2}\theta\,\sqrt{1-L^2 \cos^2 \theta
\,\phd^2-L^2\,\dot{\theta}^2}+L\sin^{n-1} \theta\, \phd
\right]\ .
\labell{newact}
\eeq
Now we make a Legendre transform to eliminate $\phd$ in favor of $\Pp$
to produce a Routhian $\cR_n(\Pp,\dot{\theta},\theta)$.
Next we analytically continue to
euclidean time, $\tau\rightarrow iz$, which yields
\beqa
\cR^E_{n}&=&{N\over L}\sqrt{1+L^2(\theta')^2}V(\theta)\ ,\\
V(\theta)&\equiv&\sqrt{p^2+\tan^2\theta(p-\sin^{n-3}\theta)^2}\ ,
\eeqa
where $\theta'=\partial_z\theta$. Rather than work with the second
order euclidean equations of motion,
it is easier to find the instanton solution
by evaluating the corresponding conserved ``energy'',
\beq
\cH^E={\delta \cR^E_{n}\over\delta\theta'}\theta'-\cR^E_{n}
=-{N\over L}{V(\theta)\over\sqrt{1+L^2(\theta')^2}}\ .
\eeq
At the extrema, $\sin\theta=0$ and $\sin^{n-3}\theta=p$,
which will correspond to
the instanton end-points, this evaluates to
$\cH^E=-Np/L$. Hence we simply need to solve
\beq
{V(\theta)\over\sqrt{1+L^2(\theta')^2}}=p\ ,
\labell{eeom}
\eeq
which has the remarkably simple solutions
\beq
\sin^{m-3}\theta_{\pm}(z)={p\over 1+e^{\pm(z-z_0)/L}}\ .
\eeq
The two solutions correspond to the instanton (+) describing a transition 
from
the expanded \nt-brane (at $z\rightarrow-\infty$) to the point-like 
configuration at
($z\rightarrow\infty$); and the anti-instanton ($-$) describing the opposite
transition. The constant $z_0$ is an integration constant giving the
instanton position in euclidean time. See Figure~\ref{fig4} for a plot
of the solution.

Next one would like to calculate the euclidean action. Because $\cR^E_n$ 
remains finite
when evaluated at either of the extrema, evaluating $\int dz \cR^E_n$ for 
the instanton
configuration would yield an infinite result. Instead, the relevant quantity 
to evaluate is
\beq
\cS^E_n=\int_{-\infty}^\infty dz\,\left(\cR^E_n(\theta_{\rm instanton})
-\cR^E_n(\theta_{\rm end-point})\right)\ ,
\labell{eact}
\eeq
which will give a measure of the degree of tunneling.
Using eq.~(\ref{eeom}), we find
\beq
\Delta \cR^E_n\equiv \cR^E_n(\theta)-\cR^E_n(\theta=0)
={N\over Lp}\left[V(\theta)^2-p^2\right],
\eeq
and the total action evaluates to
\beq
\cS^E_n=N\int_{0}^{\arcsin(p^{1/(n-3)})}d\theta
\,\sqrt{V(\theta)^2-p^2}
=N\int_{0}^{\arcsin(p^{1/(n-3)})}d\theta\,
\tan\theta(p-\sin^{n-3}\theta)\ .\;\labell{ninstact}
\eeq
We can evaluate this explicitly for the cases of interest:
\beqa
\cS^E_{n=4}&=&{N\over 2}\left[(1-p)\ln(1-p)-(1+p)\ln(1+p)+2p\right]\ ,\\
\cS^E_{n=5}&=&{N\over 2}\left[(1-p)\ln(1-p)+p\right]\ ,\\
\cS^E_{n=7}&=&{N\over 2}\left[(1-p)\ln(1-\sqrt{p})+\sqrt{p}+p/2\right]\ .
\eeqa
In all the cases these are monotonically increasing functions of $p$,
with finite limits as $p\rightarrow 1$.

\begin{figure}[htb]
\centerline{\includegraphics[width=6.5cm]{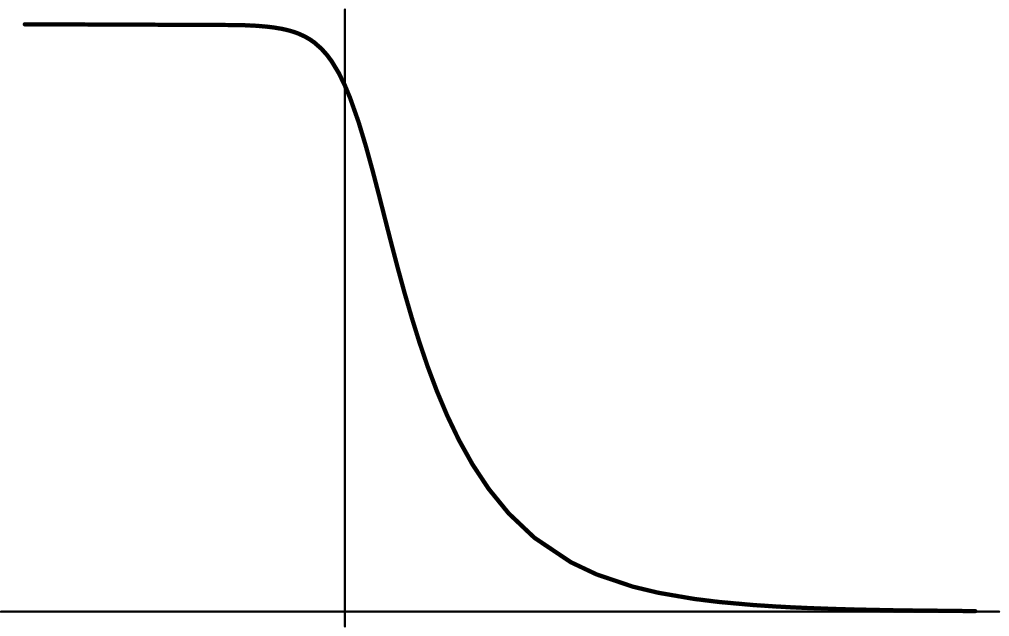}}
\begin{picture}(0,0)
\put(52,45){\small Expanded}
\put(70,8){\small $z{=}z_0$}
\put(100,14){\small Zero-size}
\put(118,8){\small $z$}
\put(76,50){\small $r_+(z)$}
\end{picture}
\vspace{-1cm}
{
\caption{Instanton transition from the expanded \mt-brane on AdS$_m$
to the collapsed one (plot made for $m=7,\ \hp=1$).}\label{fig5}
}
\end{figure}

We can repeat the analysis to find instantons for the \mt-brane which 
expands
on the AdS space. One begins in this case by
allowing a time dependent radius, $r(\tau)$. In terms of the scaled
radius, $\hr=r/\tL$, the euclidean Routhian is
\beq
\cR^E_{m}={\tN\over L}\left[\sqrt{1+{\tL^2(\hr')^2\over(1+\hr^2)^2}}
\sqrt{(1+\hr^2)(\hp^2+\hr^{2m-4})}-\hr^{m-1}\right]\ .
\eeq
Proceeding as above, we derive a first order equation for the instanton,
which may be written as
\beq
\tL\hr'=\pm{\hr(1+\hr^2)(\hp-\hr^{m-3})\over \hp+\hr^{m-1}}\ .
\label{insteq2}
\eeq
In the cases of interest, it is relatively straightforward to
integrate this equation to yield the solution
\beq
e^{\pm2(z-z_0)/L}=(1+\hr^{-2})^{m-3\over2}\left(\hp-\hr^{m-3}\right)
\ .\labell{minst}
\eeq
These provide implicit solutions (which can be solved
explicitly) for the desired instantons, evolving between the
expanded \mt-brane and the point-like one, see Figure~\ref{fig5}.

In evaluating the euclidean action, we use eq.~(\ref{insteq2}) to find
\beq
\Delta \cR^E_m\equiv \cR^E_m(\theta)-\cR^E_m(\theta=0)
={\tN\over L}{\hr^2(\hp-\hr^{m-3})^2\over \hp+\hr^{m-1}}\ ,
\eeq
and the total action becomes
\beq
\cS^E_m={\tN\tL\over L}\int_0^{p^{1/(m-3)}}
\!\!\!\!\!d\hr\ \hr\,{\hp-\hr^{m-3}
\over 1+\hr^2}
={\tN\tL\over2L}\int_0^{p^{2/(m-3)}}
\!\!\!du\,{p-u^{(m-3)/2}\over1+u}\ .\labell{minstact}
\eeq
Explicitly for AdS$_4$, AdS$_5$, and AdS$_7$, we find
\beqa
\cS^E_{m=4}&=&{\sqrt{2N}\over4}\left[\hp\ln(1+\hp^2)+2\arctan\hp-2\hp\right]\ 
,\\
\cS^E_{m=5}&=&{N\over2}\left[(1+\hp)\ln(1+\hp)-\hp\right]\ ,\\
\cS^E_{m=7}&=&N^2\left[(\hp-1)\ln(1+\sqrt{\hp})+\sqrt{\hp}-\hp/2\right]\ .
\eeqa
It is no coincidence that these expressions resemble those following
from eq.~(\ref{ninstact}) --- introducing $u=\sin^2\theta$ in the
latter produces an integrand identical to the one in
eq.~(\ref{minstact}), except
that the denominator is replaced by $1-u$.

\section{SUSY versus Goliath} \labels{susy}
\subsection{  Supersymmetry for Giants}
In this subsection we make a detailed analysis of the supersymmetries
preserved by the expanded M2-brane configuration
on AdS$_7\times \S^4$ described above, and we confirm
the expectation that (for $p\le 1$) it preserves the same
supersymmetries as the corresponding `point-like' supergraviton.
Our conventions here largely follow those of the review by Duff \cite{duff}.

Eleven-dimensional  supergravity is described by the
vielbein ${E_M}^A(X)$, the
gravitino $\Psi_M(X)$, and the three-form potential $A^{(3)}_{MNP}(X)$ with  
field
strength $F^{(4)}_{MNPQ}$. Also, in a superspace formulation, one  
introduces a
six-form  $A^{(6)}_{MNPQRS}$ whose field strength is dual to  
$F^{(4)}_{MNPQ}$.
For the AdS$_7\times \S^4$ compactification, we are
considering the Freund-Rubin ansatz \cite{freund} with the four-form 
$F^{(4)}=
{3/L}\,\varepsilon(\S^4)$, where $\varepsilon(\S^4)$ is the volume form on
the four-sphere.
The spinors  are Majorana, and the gamma-matrices satisfy $\{ \Ga^A, 
\Ga^B\}=2
\eta_{AB}$ with $\eta^{AB} = diag(-,+, \cdots ,++)$.
We also adopt the standard notation: $\Gamma^{AB\cdots C}=\Gamma^{[A}
\Gamma^B\cdots\Gamma^{C]}$. (We only write eleven-dimensional
gamma-matrices in the following.)
The SUSY transformation of the gravitino is
\beq
\delta \Psi_M = \tilde{D}_M \eps - \frac{1}{288}({\Gamma_M}^{PQRS}-8
{\delta_M}^P \Gamma^{QRS})F_{PQRS}\eps\ ,
\labell{varox}
\eeq
where $\tilde{D}_M$ is a supercovariant derivative containing the usual
connection augmented by gravitino-dependent terms.
A bosonic configuration ($\Psi_M=0$) will respect the supersymmetry for 
parameters $\eps$
such that $\delta \Psi_M =0$ (since the bosonic fields vary into the 
gravitino).

For the AdS$_7\times \S^4$ background, $\Psi_M=0$ and so the supercovariant
derivative reduces to an ordinary  covariant derivative in eq.~\reef{varox}.
Thus in this background, the Killing spinor equations ($\delta \Psi_M =0$)
become
\beqa
&&({D}_m -\frac{1}{2L}\gamma \Gamma_m)\eps=0\ ; \nonumber\\
&&({D}_\mu -\frac{1}{4L}\gamma \Gamma_\mu)\eps=0\ ;
\labell{death}
\eeqa
for $m$ and $\mu$ on $\S^4$ and  AdS$_7$, respectively.
Implicitly here and in the following, we are using the natural orthonormal
frame pointing along the coordinate directions for the metrics given in
eqs.~\reef{adsmet} and \reef{sphmet}. Recall that
the AdS$_7$ coordinates are
$t, r, \alpha_1,\ldots,\alpha_5$, and those on $\S^4$ are 
$\theta,\phi,\chi_1,\chi_2$.
In eq.~\reef{death}, $\gamma \equiv \Gamma^{\theta \phi \chi_1 \chi_2}$.
The Killing spinors may then be written as  (see also ref. \cite{Lu})
\beq
\eps(X) =
  e^{{1\over2}\theta \gamma \Gamma^{\theta}}
e^{{1\over2}\phi \gamma \Gamma^{\phi}}
e^{-{1\over2}\chi_1\Gamma^{\chi_1 \theta}}
e^{-{1\over2}\chi_2\Gamma^{\chi_2 \chi_1}}\eta\ ,
\labell{eps}
\eeq
where
\beq
\eta (t,r, \alpha_1, \cdots \alpha_5) = e^{{1\over2}\alpha\Gamma^r
\gamma} 
e^{-\frac{t}{4L}\Gamma^t\gamma} 
e^{{1\over2}\alpha_1\Gamma^{\alpha_1r}}
e^{{1\over2}\alpha_2\Gamma^{\alpha_2\alpha_1}}
\cdots
e^{{1\over2}\alpha_5\Gamma^{\alpha_5 \alpha_4}}\eps_0\ .
\labell{eta}
\eeq
Here  $\alpha = {\rm sinh}^{-1}(r/\tilde{L})= {\rm sinh}^{-1}(r/2L)$.
Note that the `AdS gamma matrix' factors in eq.~\reef{eta} commute
with the `four-sphere gamma matrix' factors in eq.~\reef{eps}, and so
one is free to re-order these exponentials.
Finally in eq.~\reef{eta}, $\eps_0$ is an arbitrary constant
spinor, and thus one has that the
AdS$_7\times \S^4$ background is maximally supersymmetric, preserving
32 supersymmetries.

Now in curved superspace with coordinates $Z^M=( X^m, \Theta^\mu)$ and
11-dimensional supergravity described by the super-vielbein ${E_M}^A(Z)$ and
super-three-form $A^{(3)}_{ABC}(Z)$, the M2-brane action is given by
\beq
S= T_2 \int d^3 \sigma  \left[-\frac{1}{2} \sqrt{-g}g^{ij} {E_i}^a {E_j}^b
\eta_{ab}
+\frac{1}{2} \sqrt{-g} + \frac{1}{3!} \veps^{ijk}{E_i}^A{E_j}^B{E_k}^C 
A^{(3)}_{CBA}
\right]\ ,
\labell{action}
\eeq
where the pull-back is defined by
${E_i}^A =\partial_iZ^M{E_M}^A$ and $A= (a, \alpha)$ are tangent space 
indices.
The action is invariant under target-space supersymmetry and, when the
supergravity constraints are satisfied, under $\kappa$-symmetry. The bosonic
part of the action reduces to that given in eq.~\reef{actp} if we substitute
in the solution for the world-volume metric equations of motion,
\beq
g_{ij}= \prt_i X^M\,\prt_j X^M\, G_{MN}(X)\ .
\labell{metsol}
\eeq

For the brane configuration studied in the Section 2.1, we have set
the spinor coordinates $\Theta=0$, as well as having set
the gravitino field $\Psi_M=0$ in the background spacetime.
This configuration will have residual supersymmetry if the following 
combined
target-space supersymmetry and $\kappa$-symmetry transformations can be 
satisfied
for some spinor(s) $\eps (X)$:
\beqa
&&\delta \Psi_M = [{D}_M  - \frac{1}{288}({\Gamma_M}^{PQRS}-8
{\delta_M}^P \Gamma^{QRS})F_{PQRS}]\eps (X) =0\ , \labell{transfa}\\
&& \delta \Theta = \left.\eps(X)\right\vert_{M2}+(1+\Gamma)\kappa 
(\sigma)=0\ ,
\labell{transfb}
\eeqa
where
\beq
\Gamma = -\frac{1}{3!}\veps^{ijk} \prt_iX^M\prt_jX^N\prt_kX^P\Gamma_{MNP}\ .
\labell{Gamma}
\eeq
The first of these constraints \reef{transfa} are, of course, the background 
Killing
spinor equations, which have the solutions given in eqs.~\reef{eps} and 
\reef{eta}.
Note that the second equation is only evaluated (defined) on the M2-brane 
world-volume.

The usual strategy \cite{bergshoeff}, which we follow here, is to first 
construct
the Killing spinors (given above) and then enforce the second equation.
After imposing the $\kappa$-symmetry gauge choice $(1+\Gamma)\Theta=0$ and 
requiring that it
be preserved by the transformation \reef{transfb}, one finds a relation 
between $\eps$ and
$\kappa$. Using this, the second constraint \reef{transfb} is rewritten as
\beq
\Gamma \eps = \eps\ ,
\labell{cond}
\eeq
which again is only evaluated at the position of the M2-brane.

For the particular embedding of the membrane described in eqs.~\reef{stat}
and \reef{try} we find
\beqa
{\Gamma} &=&\frac{1}{\sqrt{-g}}\,\prt_\tau X^M  \prt_{\sigma_1}X^N
\prt_{\sigma_2}X^P \,{E^A}_M {E^B}_N {E^C}_P\,
{\Gamma}_{ABC} \nonumber\\
&=& \frac{1}{L^2 \sin^3\theta \sin\chi_1} [ {E^{\hat{t}}}_t
{E^{\hat{\chi}_1}}_{\chi_1}
{E^{\hat{\chi}_2}}_{\chi_2}\,{\Gamma}_{\hat{t}\hat{\chi}_1\hat{\chi}_2}
+\dot{\phi}{E^{\hat{\phi}}}_\phi {E^{\hat{\chi}_1}}_{\chi_1}
{E^{\hat{\chi}_2}}_{\chi_2}\,{\Gamma}_{\hat{\phi}\hat{\chi}_1\hat{\chi}_2 }]
\nonumber\\
&=&{-\frac{1}{\sin \theta}}[ {\Gamma}^{t\chi_1 \chi_2} -\cos\theta
\,{\Gamma}^{\phi\chi_1 \chi_2}]  \,.
\labell{gammaexpl}
\eeqa
We have employed (and then dropped) hatted indices to distinguish between
local and tangent space indices. Also as stated above, we are using the 
natural vielbein
directed along the coordinate directions, and so $E^A{}_M$ can be
read off from the metrics in eqs.~\reef{adsmet} and \reef{sphmet}.
It is straightforward to check that $ \Gamma^2=1$. Note that in going
between the second and third line in eq.~\reef{gammaexpl}, we have 
substituted
$\phd=1/L$, which is satisfied for both of the stable brane configurations.

Now we write the condition in \reef{cond} as
\beq
[ \Gamma^{t\chi_1 \chi_2} -\cos\theta {\Gamma}^{\phi
\chi_1 \chi_2} + \sin \theta ]\eps =0\ ,
\eeq
or
\beq
-\Gamma^\theta \gamma[  \Gamma^{t \phi} +\cos \theta -\sin \theta \gamma
\Gamma^\theta]\eps
= -\Gamma^\theta \gamma[  \Gamma^{t \phi} + e^{-\theta \gamma\Gamma^\theta 
}]
\eps =0\ .
\labell{finalcond}
\eeq
Substituting in the Killing spinors (\ref{eps},\ref{eta}),  we must
satisfy
\beqa
0&=& [  \Gamma^{t \phi} + e^{-\theta \gamma\Gamma^\theta }]
e^{\frac{1}{2} \theta \gamma \Gamma^{\theta}}
e^{\frac{1}{2} \phi \gamma \Gamma^{\phi}}
e^{-\frac{1}{2} \chi_1  \Gamma^{\chi_1\theta}}
e^{-\frac{1}{2} \chi_2  \Gamma^{\chi_2 \chi_1}} 
e^{{1\over2}\alpha\Gamma^r
\gamma} e^{-\frac{t}{4{L}} \Gamma^t
\gamma} \cdots  \eps_0
\nonumber\\
&=& e^{-\frac{1}{2} \theta \gamma \Gamma^{\theta}}
[\,\Gamma^{t \phi} + 1\,]
e^{\frac{1}{2} \phi \gamma \Gamma^{\phi}}
e^{-\frac{1}{2} \chi_1  \Gamma^{\chi_1\theta}}
e^{-\frac{1}{2} \chi_2  \Gamma^{\chi_2 \chi_1}} 
e^{{1\over2}\alpha\Gamma^r\gamma} 
e^{-\frac{t}{4{L}} \Gamma^t\gamma} \cdots  \eps_0\ .
\labell{finalform}
\eeqa
In the second line above, we have used the fact that $\Gamma^{t\phi}$ and
$\gamma\Gamma^\theta$ anticommute, and so
\beq
\Gamma^{t \phi}e^{\frac{1}{2} \theta \gamma \Gamma^\theta} =e^{-\frac{1}{2}
\theta \gamma \Gamma^\theta}\Gamma^{t \phi}\ .
\eeq
Now the remaining exponentials obviously commute with $\Gamma^{t \phi}$, 
except for the one
containing $\Gamma^r\gamma$. However, the test brane configurations sit at
$r=\tilde{L}{\rm sinh}\alpha=0$, and
so this factor reduces to the identity 
when evaluated on the M2-brane world-volume.
Thus, the final condition
\reef{finalform} amounts to imposing
\beq
(\Gamma^{t \phi}+1)\eps_0 =0 ~.
\labell{endresult}
\eeq
That is, supersymmetry will be preserved with the constant spinors $\eps_0$ 
satisfying
this condition.
We note that  $(\Gamma^{t \phi})^2 =1$ and tr$(\Gamma^{t \phi}) =0$ so that
the spherical M2-brane preserves half of the supersymmetries of
eleven-dimensional supergravity.

Note that the equation $\phd=1/L$, which was used in eq.~\reef{gammaexpl},
was crucial to this derivation of the final constraint \reef{endresult}.
Thus the supersymmetry is preserved by precisely the two M2-brane solutions
sitting at the stable minima of the potential $\cH_{n=4}$ given in 
eq.~\reef{ham2}.
So both the expanded brane and the point-like state are BPS 
configurations,
and they both preserve precisely the same supersymmetries.
These configurations are supposed to describe a massless particle
(\eg a graviton) moving along the $\phi$ direction in the eleven-dimensional 
spacetime.
The projection \reef{endresult}
is in accord with what one might have expected by examining
the supersymmetry of gravitational waves propagating in flat space ---
see, for example, ref.~\cite{wave}.

\subsection{Supersymmetry for `Dual' Giants}

We consider now the situation of an M5-brane moving on  AdS$_7 \times \S^4$,
where the brane expands into the AdS$_7$ part of the space.
We shall show that the expanded M5-brane preserves precisely the same
supersymmetries as the M2-brane configurations.

The residual supersymmetries for a purely bosonic M5-brane configuration
must again satisfy two constraints, one being that for the spacetime Killing 
spinors
\reef{transfa} and the other involving combined supersymmetry and 
$\kappa$-symmetry
transformations on the world-volume. The full analysis is slightly more
complicated for the M5-brane because of the self-dual three form
in the world-volume theory --- see, for example, ref.~\cite{5brane,duff}.
However, when the self-dual three-form on the world-volume vanishes, as it 
does
for the configurations of interest, the final result is that we must find
Killing spinors that now satisfy the constraint
\beq
\Gamma \eps =\eps\ ,
\labell{louse}
\eeq
where
\beq
\Gamma=-\frac{1}{6!}\veps^{i_1\cdots i_6} \prt_{i_1}X^{M_1}
\cdots\prt_{i_6}X^{M_6}\Gamma_{M_1\cdots M_6}\ .
\labell{lousy}
\eeq

Given the embedding in eqs.~\reef{stat2} and \reef{try2} with $m=7$,
we obtain from the corresponding induced metric
\beq
\sqrt{-g}= \sqrt{1+\frac{r^2}{4L^2}- L^2 \phd^2}\
r^5 \sin^4\alpha_1 \sin
^3\alpha_2\cdots \sin \alpha_4\ ,
\eeq
using $\tilde{L}=2L$ for the present background.
Correspondingly, we find
\beq
\Ga =  \frac{1}{\sqrt{1+\frac{r^2}{4L^2} -L^2 \phd^2}}
\left[ - \sqrt{1+\frac{r^2}{4L^2}}\Gamma^{t \alpha_1
\cdots \alpha_5} +L
\dot{\phi}\, \Ga^{\phi \alpha_1
\cdots \alpha_5 }\right]\ .
\eeq
With $\phd =1/L$, the
condition \reef{louse}  can be written as
\beq
\left[ \frac{r}{2L} +\sqrt{1+\frac{r^2}{4L^2}}  \Gamma^{t\alpha_1
\cdots \alpha_5} -\Gamma^{\phi  \alpha_1 \cdots \alpha_5}\right]
\eps=0
\labell{condi}
\eeq
At this point, we note that the product of all of the eleven dimensional 
gamma matrices
yields the identity, \ie $\Ga^{tr\al_1\cdots\al_5\theta\phi\chi_1\chi_2}=1$.
This allows us to write: $\Gamma^{t\alpha_1 \cdots \alpha_5}= - \Gamma^r 
\gamma$
and  $\Gamma^{\phi\alpha_1 \cdots \alpha_5}=  \Gamma^r \gamma\,\Ga^{t\phi}$
where as above $\gamma =\Gamma^{\theta \phi \chi_1 \chi_2}$.
It is also useful to introduce ${\rm sinh}\alpha = \frac{r}{2L}$ as
in eq.~\reef{eta}. Then the constraint \reef{condi} reduces to
\beq
\left[ e^{-\alpha \Gamma^r \gamma} +\Gamma^{t \phi}\right] \eps =0\ .
\eeq
Now given the Killing spinors in eqs.~\reef{eps} and \reef{eta},
one can verify that this condition amounts to
\beq
(\Gamma^{t \phi}+1)\eps_0=0
\eeq
precisely as in the previous section. In deriving this final result, one 
uses that
$\Gamma^{t\phi}$ and $\Ga^r\gamma$ anticommute and that the brane 
configurations
of interest sit at $\theta=0$.

\subsection{Gravitons on ${\rm AdS}_4\times\S^7$} \labels{five}

In the following two subsections we repeat the above analysis for two
other cases of interest, M5-branes expanding on $\S^7$ and M2-branes
expanding on AdS$_4$. For the present compactification the Freund-Rubin 
ansatz
\cite{freund} sets $F^{(4)}= \frac{6}{L}\varepsilon({\rm AdS}_4)$
where $\varepsilon({\rm AdS}_4)$ is the volume form on AdS$_4$.
The Killing spinor conditions now become
\beqa
&&({D}_m -\frac{1}{2L}\gamma \Gamma_m )\eps =0 \nonumber\\
&&({D}_\mu + \frac{1}{2 \tL} \Gamma_\mu \gamma) \eps =0
\labell{neweq}
\eeqa
on  S$^7$ and  AdS$_4$, respectively, with the new definition
$\gamma \equiv \Gamma^{tr\alpha_1\alpha_2}$.
The solutions have a similar form to those found previously in 
eqs.~\reef{eps}
and \reef{eta}:
\beqa
\eps &=& e^{\frac{1}{2} \theta \gamma \Gamma^\theta} e^{ \frac{1}{2} \phi
\gamma \Gamma^\phi} e^{- \frac{1}{2} \chi_1 \Gamma^{\chi_1 \theta}}
\prod_{j=2}^{5}{} e^{- \frac{1}{2} \chi_j \Gamma^{\chi_{j} \chi_{j-1}}}
\nonumber\\
&&\quad\times e^{-\frac{\alpha}{2}\Gamma^r\gamma} 
e^{\frac{t}{L}\Gamma^t\gamma}
e^{-\frac{\alpha_1}{2}\Gamma^{\alpha_1r}}
e^{-\frac{\alpha_2}{2}\Gamma^{\alpha_2\alpha_1 }}\eps_0\ .
\labell{kspinors}
\eeqa
where $\alpha$ is now defined by ${\rm sinh}\alpha = 2r/L$.
Again $\eps_0$ is an arbitrary 32-component constant spinor, and so
AdS$_4\times\S^7$ provides another maximally supersymmetric background for
the eleven-dimensional supergravity.

\subsubsection{Giant Gravitons in $\S^7$}

The analysis of the supersymmetry
is similar to the case of the M2-branes on $\S^4$.
The M5-branes are embedded as in eqs.~\reef{stat} and \reef{try} with
$n=7$. The induced metric yields
$\sqrt{- g}= L^5 \sin^6 \theta \sin^4 \chi_1 \cdots \sin \chi_4$
when evaluated for $\dot{\phi}= 1/L$. Then using  appropriate
expressions for the vielbein, eq.~\reef{lousy} yields
\beq
\Gamma = -\frac{1}{\sin \theta}(\Gamma^{t\chi_1 \cdots \chi_5}-
\cos \theta \Gamma^{\phi\chi_1 \cdots \chi_5} )\ .
\eeq
In this case, we impose the supersymmetry condition   $\Gamma \eps = \eps$
which can be manipulated to the form
\beq
(\Gamma^{t \phi} + e^{- \theta \gamma \Gamma^\theta})\eps =0\ .
\eeq
It is satisfied by the Killing spinors \reef{kspinors}
provided
\beq
(\Gamma^{t \phi} +1) \eps_0 =0 \ .
\eeq

\subsubsection{Giant Gravitons in ${\rm AdS}_4$}

The counterpart of the previous case is that of the M2-brane
expanding in AdS$_4$. The embedding is now  as in eqs.~\reef{stat2}
and \reef{try2}, and the induced metric leads to $\sqrt{-g}= R^3/ \tilde{L}
\sin^2\alpha_1 \sin \alpha_2$.
The supersymmetry  condition   is now $\Gamma \eps = - \eps$
as required for anti-M2-branes --- compare to eq.~\reef{cond} in Section~3.1.
The constraint becomes
\beq
\left[\sqrt{1+\frac{4r^2}{L^2}}\Ga^{t\alpha_1 \alpha_2} - \Ga^{\phi \alpha_1
\alpha_2} -   \frac{2r}{L}\right]\eps=0\ ,
\eeq
substituting $\tilde{L} = L/2$. Using ${\rm sinh}\alpha = 2r/L$, we have
\beq
\Ga^{t \alpha_1 \alpha_2}[e^{-\alpha \Ga^{t \alpha_1 \alpha_2} }
+\Ga^{t \phi}]\eps= 0\ .
\eeq
Now using $\Ga^{t \alpha_1 \alpha_2}= -\Ga^r \gamma$, the condition
becomes
\beq
[e^{\alpha \Ga^{r \gamma} }+\Ga^{t \phi}]\eps=0\ ,
\eeq
and using the explicit form of the  Killing spinors in eq.~\reef{kspinors},
it reduces once
more to the same condition found above,
\beq
(\Gamma^{t \phi} +1) \eps_0 =0 \ .
\eeq

Some comments on the supersymmetry projections used here are in order.
Note that for the spherical M2-branes on $\S^4$, the supersymmetry 
condition was $\Gamma \eps = \eps$ in section 3.1, while for the M2-branes
expanding into AdS$_4$,
we imposed the condition $\Gamma \eps = -\eps$ above. These choices
are in agreement with the calculations in Section~2. Recall that, as remarked after
eq.~\reef{clown}, with our conventions while we have M2-branes expanding on $\S^4$, 
the `dual' giant gravitons expanding in AdS$_4$ are anti-M2-branes.
With regard to the residual  supersymmetries,
the difference between branes and anti-branes amounts to reversing a 
projection
in the relevant $\kappa$-symmetry transformations. This in turn results in 
reversing the sign of the final constraint imposed on the Killing spinors. Hence we 
find the opposite supersymmetry constraints are satisfied for the expanded 
M2-branes on $\S^4$ here and those on AdS$_4$ in Section~3.1. In contrast,
the projection imposed for the expanding M5-branes took the form $\Gamma\eps=\eps$ for
both AdS$_7$ and $\S^7$. This is also in accord with the results of Section~2,
where we found that both cases corresponded to anti-M5-branes within our conventions
--- see the footnote after eq.~\reef{lag2} --- and hence the same supersymmetry constraint
applies in both cases.

\subsection{Gravitons on ${\rm AdS}_5\times \S^5$} \labels{three}

Finally, we consider D3-branes propagating in a 10-dimensional
type IIb background compactified on AdS$_5 \times \S^5$.
Ten-dimensional, Type IIb supergravity is described by the vielbein, a 
complex
Weyl gravitino, a real four-form  $A^{(4)}_{MNPQ}$ with {\em self-dual} 
field
strength  $F^{(5)}_{MNPQR}$, a complex two-form $A^{(2)}_{MN}$, a complex 
spinor $\Lambda$
and a complex scalar $\Phi$. In the AdS$_5 \times \S^5$ background, we
have $\Phi=  A^{(2)}_{MN} = \Lambda =\Psi_M=0$.
The five-form field strength is
  $F= \frac{4}{L}[\varepsilon({\rm AdS}_5)+\varepsilon( \S^5)]$.
Given that the scalar and two-form vanish in this background, the
supersymmetry variation of the complex spinor automatically vanishes,
$\delta\Lambda=0$. Hence to examine the background supersymmetries, the
only nontrivial variation which needs to be considered is that of the 
gravitino.
For the given background, the variation of the
gravitino takes the form
\beq
\delta \Psi_M = {D}_M \eps - \frac{i}{480}{\Gamma_M}^{PQRST}F_{PQRST}\eps\ .
\eeq
Demanding $\delta\Psi_M=0$ leads to the Killing spinor equations
\beq
{D}_M \eps -\frac{i}{4}(\Gamma^{tr \alpha_1 \alpha_2 \alpha_3}
+ \Gamma^{\theta \phi
\chi_1  \chi_2  \chi_3})\Gamma_M \eps =0\ .
\eeq
We recall that here $ \eps$ is a complex Weyl spinor satisfying $\Gamma^{11}
\eps = \eps$ where  $\Gamma^{11}=
\Gamma^{tr\alpha_1 \alpha_2 \alpha_3\theta \phi\chi_1  \chi_2  \chi_3}$.
(Our gamma matrices  are now ten-dimensional.)
The Killing spinor condition can be rewritten then
as
\beqa
&&{D}_\mu \eps -\frac{i}{2}\gamma_{AdS}\Gamma_\mu \eps =0
~~~,~~~~\gamma_{AdS}=\Gamma^{t r\alpha_1 \alpha_2 \alpha_3}
\equiv \gamma\ , \nonumber\\
&&{D}_m \eps -\frac{i}{2}\gamma_{5}
\Gamma_m \eps =0 ~~~~~,~~~~\gamma_{_5}
=\Gamma^{ \theta \phi\chi_1  \chi_2  \chi_3}\ ,
\eeqa
on AdS$_5$ and $ \S^5$, respectively.
The Killing spinor solutions are now
\beqa
\eps &=& e^{\frac{i}{2} \theta \gamma_5 \Gamma^\theta} e^{ \frac{i}{2} \phi
\gamma_5 \Gamma^\phi} 
e^{- \frac{i}{2} \chi_1 \Gamma^{\chi_{1}\theta}}
e^{- \frac{i}{2} \chi_2 \Gamma^{\chi_{2}\chi_1}}
e^{- \frac{i}{2} \chi_3 \Gamma^{\chi_{3}\chi_{2}}}
\nonumber\\
&&\quad\times e^{i\frac{\alpha}{2}\Gamma^r\gamma} 
e^{-i\frac{t}{2L}\Gamma^t\gamma}
e^{\frac{\alpha_1}{2}\Gamma^{\alpha_1r}}
e^{\frac{\alpha_2}{2}\Gamma^{\alpha_2\alpha_1}}
e^{\frac{\alpha_3}{2}\Gamma^{\alpha_3\alpha_2} }
\eps_0\ ,
\labell{killme}
\eeqa
where $\sinh\alpha = r/L$. Hence we have a maximally 
supersymmetric
solution of the type IIb supergravity equations with 32 residual 
supersymmetries.

\subsubsection{Giant gravitons in $\S^5$}

Supersymmetric world-volume actions have been constructed for all type II
D$p$-branes in a general supergravity background \cite{3brane}, but we will
not elaborate on any of the details here. The analysis for determining the
residual supersymmetries of a D-brane configuration is similar to that
for their M-theory cousins. For our D3-brane configurations (in
which the world-volume gauge fields vanish), residual supersymmetries
are again determined by imposing a constraint on the background Killing 
spinors
of the form $\Ga \eps=\pm\eps$, now using the matrix
\beq
\Gamma=-\frac{i}{5!}\veps^{i_1\cdots i_5} \prt_{i_1}X^{M_1}
\cdots\prt_{i_5}X^{M_5}\Gamma_{M_1\cdots M_5}\ .
\labell{lousya}
\eeq
For the giant gravitons on $\S^5$, we consider the embedding of a
D3-brane as in eqs.~\reef{stat} and \reef{try}.
With this configuration, one finds
\beq
\Gamma = -\frac{i}{\sin \theta}(\Gamma^{t \chi_1 \chi_2\chi_3} -
\cos \theta  \Gamma^{\phi \chi_1 \chi_2\chi_3} )\ ,
\eeq
and the condition $\Gamma\eps=\eps$
becomes, after pulling out a factor of $\Gamma^{\phi \chi_1 \chi_2\chi_3}$,
\beq
(e^{-i \theta \gamma_5 \Gamma^{\theta}} + \Gamma^{t \phi})\eps=0\ .
\eeq
It is again easy to verify that the various exponentials in the
Killing spinors \reef{killme} can be pulled through so as to reduce the 
condition
to
\beq
(\Gamma^{t\phi}+1)\eps_0=0\ .
\eeq

\subsubsection{Giant gravitons in ${\rm AdS}_5$}

We consider now  the embedding as in eqs.~\reef{stat2} and \reef{try2}.
In the present case, the matrix $\Gamma$ takes the form
\beq
\Gamma= \frac{i}{{\rm sinh}\alpha}[ {\rm cosh}\alpha \Gamma^{t \alpha_1 
\alpha_2
\alpha_3} -  \Gamma^{\phi \alpha_1 \alpha_2
\alpha_3}]\ .
\eeq
Now we impose the condition for preservation of supersymmetry, 
$\Gamma\eps=-\eps$,
which corresponds to anti-D3-branes in our conventions.\footnote{By the conventions
of Section~2, this is again a case where branes expand in $\S^5$ but anti-branes
expand in AdS$_5$.}
This equation can be written as
\beq
\left[e^{-i \alpha \Gamma^r \gamma}+ \Gamma^{t \phi}\right]\eps =0
\eeq
with the same conclusion concerning the preservation of
supersymmetry, namely, we project the constant spinors with
\beq
(\Gamma^{t \phi}+1)\eps_0=0\ .
\eeq

\section{Discussion} \labels{discuss}

For a given background $\AdS_m\times\S^n$,
we have identified three different test-brane configurations carrying
angular momentum $\Pp$, all of which have the same energy and preserve
precisely the same supersymmetries. These configurations are
the giant graviton of ref.~\cite{giant} consisting of a spherical
\nt-brane expanding out into the $n$-sphere, a `dual' giant
graviton consisting of a spherical \mt-brane which expands out into
the AdS space, and the point-like brane at the origin.
The new configuration uncovered here is the `dual' giant graviton
which expands into the AdS space. Our motivation for looking for these
configurations came from the analysis of ref.~\cite{dielec}.
There it was shown that in type IIa superstring theory, a collection of 
D0-branes
in an {\it electric} four-form field strength will expand into a spherical 
D2-D0-brane
bound state. This situation would lift to M-theory as a spherical M2-brane
carrying momentum in a direction orthogonal to an electric four-form. The 
latter is
then essentially the `dual' giant graviton in the AdS$_4\times\S^7$ 
background.

The authors of ref.~\cite{giant} provided a simple mechanical model 
\cite{bar}
of an electric
dipole, consisting of two separated charges held together by a simple linear
restoring force and moving in a magnetic field, to provide some intuition 
for their
expanding brane calculations. One might wonder if a similar mechanical model 
might describe
the expansion of the dual branes induced by an electric fields. At first 
sight, the
the answer appears to be no. If the dipole is oriented along the external 
electric field $E$,
it is naturally extended to some equilibrium length $L$, but this length is 
unaffected by
motion of the dipole transverse to the field. Actually this discussion would 
only apply for
nonrelativistic motions. If the transverse velocity $v$ is relativistic, one 
finds there is
a noticeable effect. In this case, to calculate the effect to the restoring 
force, we
boost to the rest frame of the dipole. In this frame, however the electric 
field has
increased to $E'=\gamma E$, and so the extension of the dipole is increased 
to $L'$.
Upon boosting back to the original frame, the extension being transverse
to the boost is unaffected and
so the dipole appears to have increased in length. For small velocities, the 
variation
is small, \ie $\Delta L\propto v^2$, in contrast to the linear extension in 
a magnetic
field, \ie $\Delta L\propto v$. However, for large velocities $v\simeq c$, 
the extension
is essentially proportional to the momentum, \ie $\Delta L\propto |p|$.
Of course, this is only a crude analogy for the physics of expanding branes.
However, one observation is that reversing the dipole's
velocity in the magnetic field reverses the orientation of the stretched dipole,
but leaves it unchanged in the case of the electric field. In our brane analysis,
this corresponds to the fact that the energy \reef{green2} for the `dual' giant gravitons
is even in the angular momentum, but that in eq.~\reef{ham2} for the giant gravitons
is not. This observation is also related to the appearance of expanding branes
in some cases and expanding anti-branes in others.   
Perhaps the main lesson to draw from
this model is the reminder that the various forces acting on the elements of the
spherical branes transform in different ways under Lorentz boosts, which at least
for certain configurations allows
the expansion to take place when the branes are set in motion. From this 
point of view,
this is reminiscent of the imbalance of forces arising in the scattering of 
supersymmetric
solitons \cite{scatsol} or branes \cite{scatbra}. It would be interesting to 
investigate
if these observations have any implications for theories with
space-time noncommutativity \cite{spaced}.

The brane configurations all preserve one half of the 32 supersymmetries of 
the
background AdS$_m\times\S^n$ spacetime. The 16 supersymmetry transformations 
satisfying
the `wrong' projection, \ie $(\Ga^{t\phi}-1)\eps_0=0$, would leave the 
background
spacetime invariant but generate fermionic variations of the world-volume 
fields,
which at the same time would leave the energy invariant. Of course, the 
equations
of motion eliminate half of these to leave 8 fermionic zero-modes in each of 
the
bosonic configurations studied here. These zero-modes are regarded as 
operators
acting on a quantum space of states \cite{zero}, which then build up for 
each
bosonic configuration the full $2^8=256$ states of the supergraviton 
multiplet,
as usual.

In general, working with the test-brane action will be problematic for the 
point-like
configurations. Further, beyond the technical difficulties, we might expect 
that
extra stringy or M-theoretic corrections to the dynamics, \ie the test-brane
description will breakdown. However, the supersymmetry of these
configurations may be sufficient to protect the existence of these states
in the full theory.

Note that in counting the candidate supergraviton states, we refer to a 
single
point-like state. That is, we do not distinguish the collapsed \mt-brane
from a collapsed \nt-brane. This is motivated by the Matrix theory
description of M-theory in light-cone gauge \cite{matrix}. There in 
principle one can
represent various
different branes with different geometries using noncommutative geometry,
see \eg \cite{wati}. However, when these geometries shrink to zero size, one
is left with the same state, \ie
the same set of commuting matrices, independent of the original geometry.
Even though we do not have a fully covariant formulation of M-theory, it 
seems
natural to assume that it will still inherit this aspect of Matrix theory,
and so the different collapsed branes in our calculations should correspond 
to the
same point-like state.

In any event, it seems that we have an excess of potential states to
describe the supergravitons carrying angular momentum. A natural question
to ask is: are there even 
more expanded brane configurations with the same 
quantum
numbers? Our attempts in that direction have failed to produced 
any
additional configurations. 
The extra states beyond the original giant gravitons
already present a problem, since these branes persist as
stable configurations above the desired angular momentum bound of
$P_\phi\le N$. Certainly the nature of the supersymmetric states
carrying angular momentum changes
as $\Pp$ increases past this bound, since the expanded
branes of S$^n$ can no longer contribute. From this point of view, the 
behavior
is reminiscent of the `long strings' on AdS$_3$ \cite{long}.
In this case, macroscopic strings at infinity become relevant in describing
the spectrum of states with scaling dimension above a certain bound.

However, if we are to interpret the stringy exclusion principle as saying
there are no supersymmetric single particle states with angular momentum
$P_\phi> N$, then the extra states seem to jeopardize the proposed 
explanation of the
stringy exclusion principle in terms of expanding branes.
We would like to suggest one possible way in which stringy exclusion
principle might still be realized even though there are additional
brane configurations beyond the giant graviton states of ref.~\cite{giant}.
First, one would expect that the three candidate states will mix
quantum-mechanically. Certainly we have shown that there are
instantons allowing for tunneling between either of the expanded
brane configurations and the point-like states. It would be interesting to
find instantons mediating tunneling directly between the giant gravitons
and their `dual' cousins. However, it seems this would be beyond the scope
of the test-brane framework used here. In any event,
one might think that the various the three states mix and produce a unique
ground state representing the true graviton. 

\begin{figure}[htb]
\centerline{\includegraphics[width=6.5cm]{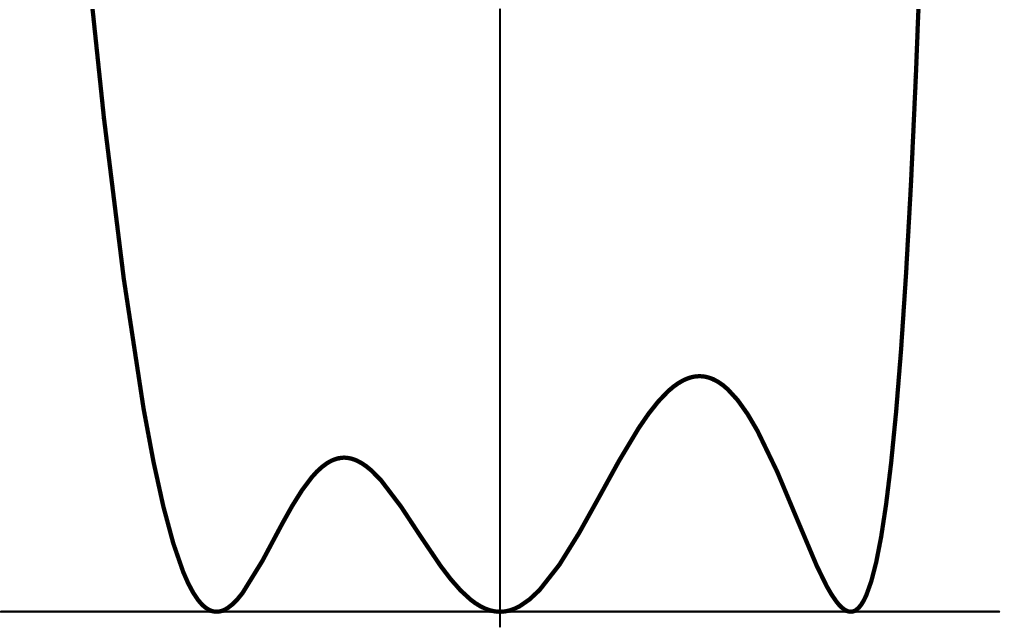}
\hspace{1.5cm}
\includegraphics[width=6.5cm]{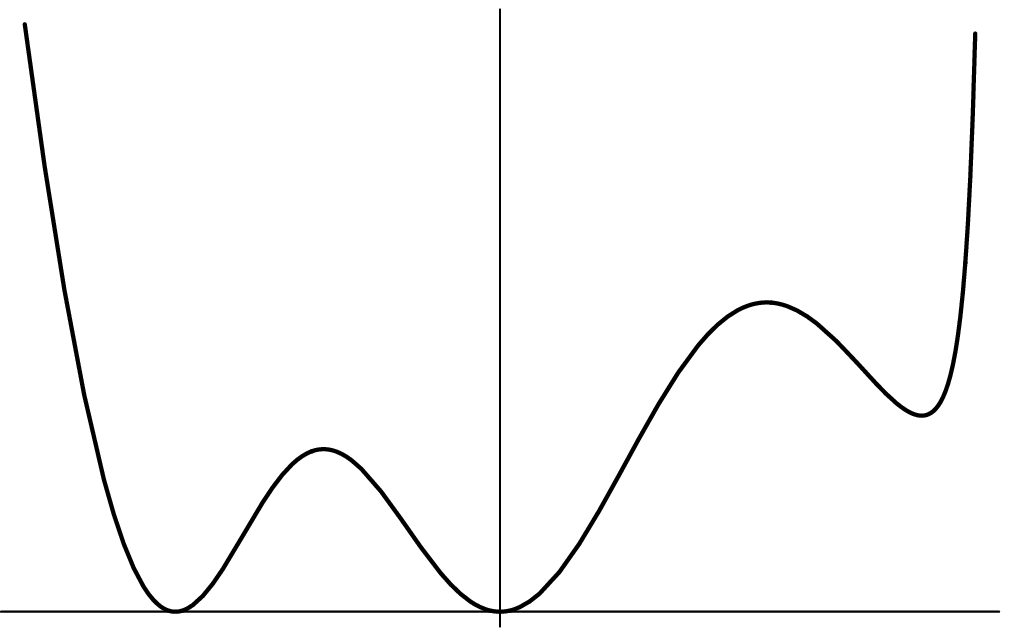}}
\begin{picture}(0,0)
\put(45,50){\small Energy}
\put(21,8){\small AdS$_m$}
\put(36,8){\small Zero-size}
\put(64.5,8){\small S$^n$}
\put(50,38){\small \bf SUSY}
\put(25,38){$P_\phi\le N$}
\put(74,13){\small Radius}
\put(128,50){\small Energy}
\put(101,8){\small AdS$_m$}
\put(119,8){\small Zero-size}
\put(134,38){\small \bf No SUSY}
\put(104,38){$P_\phi>N$}
\put(157,13){\small Radius}
\end{picture}
\vspace{-.5cm}
{
\caption{Part of the moduli space of giant gravitons. For $\Pp\le N$
there are three minima: the zero-size solution, an \mt-brane expanding
into AdS$_m$, and an \nt-brane expanding into $\S^n$. For $\Pp>N$ the
latter solution disappears --- if supersymmetry is broken in this
regime, then the stringy exclusion principle still holds.}\label{fig6}
}
\end{figure}

To realize the
exclusion principle, we further postulate that for angular
momenta beyond the exclusion principle bound, $\Pp>N$, there exists no
supersymmetric ground state once the quantum mixing is taken into
account. That is, the short supersymmetry multiplets associated with
the point-like state and the `dual' giant graviton combine to give
a massive long multiplet of states. Presumably the mass above
the BPS bound is characterized by the instanton action,
$\Delta \cH\propto \exp(-\cS^E_m)$. However, below the bound,
$\Pp<N$, there are three states and so the quantum mechanical
mixing could only lift a combination of two short multiplets and
must leave a unique supersymmetric ground state (see Figure~\ref{fig6}). 
The quantum wave function for this state 
would presumably still have support at all
three of the candidate brane configurations.

This tentative realization of the stringy exclusion principle
in terms of spontaneous symmetry breaking would be analogous to that
in supersymmetric quantum
mechanics \cite{witten} in generic models involving a single
supersymmetric coordinate. In such models, there is a crucial
difference between the cases where the bosonic potential has an even
or odd number of supersymmetric minima. For an odd number there must be
a supersymmetric ground state, while for an even number supersymmetry
can be broken completely. In the current situation, we do have an odd number
(three) of minima for $\Pp\le N$ (or more precisely, three minima for
each set of quantum numbers in the 256-dimensional supergraviton
multiplet). When we exceed the bound, $\Pp>N$, one of the minima
disappears, we are left with two minima, and this is the regime in
which we might find spontaneous supersymmetry breaking.
Of course, whether or not supersymmetry is actually broken in the present
case will depend on the fermion zero-modes associated with the instanton.
We leave determining the structure of these zero-modes as a calculation for 
future work.

It is interesting that by adjusting a parameter $\Pp$ of the theory,
one of the supersymmetric ground states disappears. This is unexpected 
behavior
in that it would be impossible with a continuous parameter in a 
supersymmetric theory
\cite{witten,minus}. However,
in the present case, the angular momentum $\Pp$ must be quantized to have
integral values. This discreteness provides the loop-hole through which the
short-multiplet supersymmetric states associated with the giant graviton
is able to disappear. This behavior is reminiscent of Witten's result
in three-dimensional N=1 super-Yang-Mills theory with a Chern-Simons 
interaction
\cite{witchen}. There one
finds that supersymmetry is spontaneously broken for a finite range of
the level number, a parameter which can only take discrete values.
Since the boundary conformal field theory for M-theory on AdS$_4$
is related to three-dimensional N=8 super-Yang-Mills, it may be interesting 
to
see if there is a concrete relation between Witten's results and the 
behavior of the giant
gravitons in the AdS$_4\times\S^7$ background.

If one considers a Kaluza-Klein reduction to the AdS$_m$ space, one has
$m$-dimensional supergravity theory with an $SO(n+1)$ gauge symmetry
arising from the isometries of the internal $\S^n$. From the point of
view of this theory, the various brane configurations considered here are 
massive
BPS states carrying a certain charge under a $U(1)$ subgroup of the 
$SO(n+1)$.
One might examine the supergravity to see if there are analogous charged
black hole solutions \cite{blacc}. However, one typically finds
that these BPS configurations in AdS space correspond to naked singularities 
\cite{roman}
rather than extreme RN black holes as in flat space. It seems then that
the expansion of the giant gravitons is very closely related to stringy 
mechanism
responsible for the removal of naked singularities by the enhan\c{c}on
found in ref.~\cite{excise}. It would be interesting to investigate this
analogy more closely.

\section*{Acknowledgments}
Research by MTG was supported by NSF Grant PHY-9604587.
Research by RCM and \O T was supported by NSERC of Canada and
Fonds FCAR du Qu\'ebec. MTG and RCM would like to thank the Aspen Center for 
Physics
for hospitality during the final stage of this project.
We would like to thank N.~Constable, B.~de~Wit, J.~Distler, I.~Klebanov,
E.~Martinec, S.~Mathur,
H.~Ooguri and M.~Strassler for useful comments and conversations.
We would also like to thank the authors of ref.~\cite{aki} for notifying
us of their results.


\begin{thebibliography}{99}

\bibitem 
{giant} J.~McGreevy, L.~Susskind and N.~Toumbas,
JHEP {\bf 0006} (2000) 008 [hep-th/0003075].

\bibitem 
{juan}
J.~Maldacena,
Adv.\ Theor.\ Math.\ Phys.\  {\bf 2} (1988) 231 [hep-th/9711200].

\bibitem 
{adscft} O.~Aharony, S.~S.~Gubser, J.~Maldacena, H.~Ooguri and Y.~Oz,
Phys.\ Rept.\  {\bf 323} (2000) 183 [hep-th/9905111].

\bibitem 
{exclus} J.~Maldacena and A.~Strominger,
JHEP {\bf 9812} (1998) 005 [hep-th/9804085];\\
A.~Jevicki and S.~Ramgoolam,
JHEP {\bf 9904} (1999) 032 [hep-th/9902059];\\
P.~Ho, S.~Ramgoolam and R.~Tatar,
Nucl.\ Phys.\  {\bf B573} (2000) 364 [hep-th/9907145];\\
S.S.~Gubser, Phys.\ Rev.\  {\bf D56} (1997) 4984 [hep-th/9704195].

\bibitem 
{aki} A.~Hashimoto, S.~Hirano and N.~Itzhaki,
{\it ``Large branes in AdS and their field theory dual,''}
hep-th/0008016.

\bibitem 
{duff} M.J.~Duff,
{\it ``TASI lectures on branes, black holes and anti-de Sitter space,''}
hep-th/9912164.

\bibitem 
{freund} P.G.O.~Freund and M.A.~Rubin, Phys. Lett. {\bf B97} (1980) 233.

\bibitem 
{5brane}
I. Bandos, K. Lechner,
A. Nurmagamabetov, P. Pasti, D. Sorokin and M. Tonin, Phys. Rev. Lett. {\bf 
78}
(1997) 4332 [hep-th/9701149],
Phys. Lett. {\bf B408} (1997) 135 [hep-th/9703127];
M. Aganagic, J. Park, C. Popescu and J.H. Schwarz,
Nucl. Phys. {\bf B496} (1997) 191 [hep-th/970166];
P.S. Howe and  E. Sezgin, Phys. Lett. {\bf B390} (1997) 133
[hep-th/9607227],
Phys. Lett. {\bf B394} (1997) 62 [hep-th/9611008];
P.S. Howe,  E. Sezgin and P.C. West, Phys. Lett. {\bf B399} (1997) 49
[hep-th/9702008].

\bibitem 
{dbrane} See, for example: C.V.~Johnson,
{\it ``D-Brane primer,''} hep-th/0007170;
J.~Polchinski,
{\it ``TASI lectures on D-branes,''}
hep-th/9611050.

\bibitem 
{3brane}
M. Cederwall, A. van Gussich, B.E.W. Nilsson and A. Westerberg,  Nucl. Phys.
{\bf B490} (1997) 163, 179 [hep-th/9610148, 9611159]; E. Bergshoeff and P.K.
Townsend, Nucl. Phys. {\bf 490} (1997) 145 [hep-th/9611173].

\bibitem 
{dielec} R.C.~Myers,
JHEP {\bf 9912} (1999) 022 [hep-th/9910053].

\bibitem 
{Lu} H.\ Lu, C.N.\ Pope and J.\ Rahmfeld, J. Math. Phys. {\bf 40} (1999) 
4518
[hep-th/9805151]

\bibitem 
{bergshoeff} E.~Bergshoeff, M.J.~Duff, C.N.~Pope and E.~Sezgin,
Phys.\ Lett.\ {\bf 199B} (1987) 69.

\bibitem 
{wave} E.~A.~Bergshoeff, R.~Kallosh and T.~Ortin,
Phys.\ Rev.\  {\bf D47} (1993) 5444 [hep-th/9212030].

\bibitem 
{bar} M.M.~Sheikh Jabbari,
Phys.\ Lett.\  {\bf B455} (1999) 129 [hep-th/9901080];\\
D.~Bigatti and L.~Susskind, 
{\it ``Magnetic fields, branes and noncommutative geometry,''}
hep-th/9908056.

\bibitem 
{scatsol}{See, for example: N.S.~Manton, Phys.\ Lett.\  {\bf B110} (1982) 
54;
R.R.~Khuri, Phys.\ Lett.\  {\bf B294} (1992) 331 [hep-th/9205052].}

\bibitem 
{scatbra}{See, for example:
G.~Lifschytz, Phys.\ Lett.\  {\bf B388} (1996) 720 [hep-th/9604156];
R.R.~Khuri and R.C.~Myers, Nucl.\ Phys.\  {\bf B466} (1996) 60
[hep-th/9512061];
D.M.~Kaplan and J.~Michelson, Phys.\ Lett.\  {\bf B410} (1997) 125
[hep-th/9707021].}

\bibitem 
{spaced} N.~Seiberg, L.~Susskind and N.~Toumbas, JHEP {\bf 0006} (2000) 021
[hep-th/0005040]; JHEP {\bf 0006} (2000) 044 [hep-th/0005015];
R.~Gopakumar, J.~Maldacena, S.~Minwalla and A.~Strominger, JHEP {\bf 0006}
(2000) 036 [hep-th/0005048].

\bibitem 
{zero} R.~Jackiw and C.~Rebbi, Phys.\ Rev.\  {\bf D13} (1976) 3398;
J.A.~Harvey, hep-th/9603086;
V.~Balasubramanian, D.~Kastor, J.~Traschen and K.Z.~Win,
Phys.\ Rev.\  {\bf D59} (1999) 084007 [hep-th/9811037].

\bibitem 
{matrix} T.~Banks, W.~Fischler, S.H.~Shenker and L.~Susskind,
Phys.\ Rev.\  {\bf D55} (1997) 5112 [hep-th/9610043].

\bibitem 
{wati} J.~Castelino, S.~Lee and W.~Taylor,
Nucl.\ Phys.\  {\bf B526} (1998) 334 [hep-th/9712105];
D.~Kabat and W.~Taylor,
Adv.\ Theor.\ Math.\ Phys.\  {\bf 2} (1998) 181 [hep-th/9711078].

\bibitem 
{long} N.~Seiberg and E.~Witten,
JHEP {\bf 9904} (1999) 017 [hep-th/9903224];
J.~Maldacena and H.~Ooguri, 
{\it ``Strings in AdS(3) and SL(2,R) WZW model. I,''} hep-th/0001053.

\bibitem 
{witten} E.~Witten, Nucl.\ Phys.\ {\bf B185} (1981) 513.

\bibitem 
{minus} E.~Witten, Nucl.\ Phys.\ {\bf B202} (1982) 253.

\bibitem 
{witchen} E.~Witten, {\it ``Supersymmetric index of three-dimensional 
gauge theory,''} in
``The many faces of the superworld:
The Yuri Golfand Memorial Volume,'' ed.~M.~Shifman (World Scientific,
2000) [hep-th/9903005].

\bibitem 
{blacc} A.~Chamblin, R.~Emparan, C.V.~Johnson and R.C.~Myers,
Phys.\ Rev.\  {\bf D60} (1999) 064018 [hep-th/9902170];
M.~Cveti\v{c} and S.S.~Gubser, JHEP {\bf 9904} (1999) 024 [hep-th/9902195];
M.~Cveti\v{c} {\it et al.}, Nucl.\ Phys.\  {\bf B558} (1999) 96
[hep-th/9903214].

\bibitem 
{roman} L.J.~Romans,
Nucl.\ Phys.\  {\bf B383} (1992) 395 [hep-th/9203018];\\
L.A.~London, Nucl.\ Phys.\  {\bf B434} (1995) 709.

\bibitem 
{excise} C.V.~Johnson, A.W.~Peet and J.~Polchinski,
Phys.\ Rev.\  {\bf D61} (2000) 086001 [hep-th/9911161];
L.~J\"{a}rv and C.V.~Johnson, 
{\it ``Orientifolds, M-theory, and the ABCD's of the enhan\c{c}on,''}
hep-th/0002244;
C.V.~Johnson, 
{\it ``Enhan\c{c}ons, fuzzy spheres and multi-monopoles,''}
hep-th/0004068.


\end{thebibliography}
\end{document}